\documentclass[aps,prl,notitlepage,superscriptaddress,showpacs,twocolumn]{revtex4-2}
\usepackage{graphicx,subfigure,epsfig}
\usepackage{array,multirow}
\usepackage{dcolumn}
\usepackage{array}
\usepackage{newtxtext}
\usepackage[smallerops,vvarbb,cmbraces,subscriptcorrection]{newtxmath}
\usepackage{bm,bbm}
\usepackage{epstopdf,color,multirow}
\usepackage{hyperref}
\usepackage{footnote}
\usepackage{ulem}
\usepackage{comment}

\hypersetup{
    colorlinks=true,
    linkcolor=blue,
    citecolor=blue,
    urlcolor=blue
}

\DeclareSymbolFont{largesymbolsCM}{OMX}{cmex}{m}{n}

\let\sum\relax
\DeclareMathSymbol{\sum}{\mathop}{largesymbolsCM}{"50}

\begin{document}
\normalem

\title{Thermal states emerging from low-entanglement background in disordered spin models}

\author{Yule Ma\href{https://orcid.org/0009-0002-8217-8170}{\includegraphics[scale=0.05]{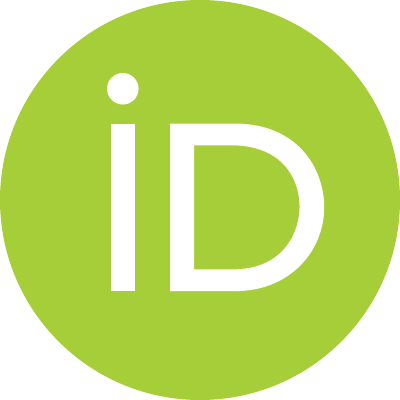}}}
\thanks{These authors contributed equally to this work.}
\affiliation{Kavli Institute for Theoretical Sciences, University of Chinese Academy of Sciences, Beijing 100190, China}

\author{Qianqian Chen\href{https://orcid.org/0000-0003-0756-408X}{\includegraphics[scale=0.05]{orcidid.pdf}}}
\thanks{These authors contributed equally to this work.}
\affiliation{Kavli Institute for Theoretical Sciences, University of Chinese Academy of Sciences, Beijing 100190, China}

\author{Mingyang Li}
\affiliation{Kavli Institute for Theoretical Sciences, University of Chinese Academy of Sciences, Beijing 100190, China}

\author{Zlatko Papi\'c\href{https://orcid.org/0000-0002-8451-2235}{\includegraphics[scale=0.05]{orcidid.pdf}}}
\email{Z.Papic@leeds.ac.uk}
\affiliation{School of Physics and Astronomy, University of Leeds, Leeds LS2 9JT, United Kingdom}

\author{Zheng Zhu\href{https://orcid.org/0000-0001-7510-9949}{\includegraphics[scale=0.05]{orcidid.pdf}}}
\email{zhuzheng@ucas.ac.cn}
\affiliation{Kavli Institute for Theoretical Sciences, University of Chinese Academy of Sciences, Beijing 100190, China}

\begin{abstract}
   Thermalization in isolated quantum systems is governed by the eigenstate thermalization hypothesis, while strong disorder can induce its breakdown via many-body localization. Here we show that disorder can also generate a narrow band of \emph{thermal} eigenstates embedded in an otherwise non-thermal spectrum. We illustrate this generic mechanism using paradigmatic spin-1 models, including Heisenberg, XY, and Affleck-Kennedy-Lieb-Tasaki (AKLT) models with several types of disorder. By analyzing their level statistics, entanglement properties and quench dynamics, we show that the disorder-induced states are genuinely thermal and we trace their origin to the null space of the disorder term in the Hamiltonian. Our results demonstrate that disorder can give rise to an unexpected coexistence of thermal and non-thermal dynamics within the same many-body spectrum.
\end{abstract}
  
\date{\today}

\maketitle

\emph{Introduction.---}A cornerstone for understanding thermalization in isolated quantum systems is the eigenstate thermalization hypothesis (ETH)~\cite{PhysRevE.50.888, Rigol2008, D'Alessio03052016, Deutsch_2018}. In recent years, considerable efforts have been devoted to exploring the mechanisms of ETH breakdown as a way of realizing long-time coherent dynamics. Typical examples of `strong' ETH violation include quantum integrable models \cite{Kinoshita2006,PhysRevLett.98.050405, PhysRevLett.103.100403, PhysRevLett.105.250401, PhysRevB.91.155123} and many-body localization (MBL) \cite{PhysRevLett.95.206603,Basko2006, PhysRevB.82.174411}, where the entire spectrum fails to thermalize. While experiments on finite-size systems have observed various signatures of MBL~\cite{doi:10.1126/science.aaa7432,smith2016many,doi:10.1126/science.aaf8834,PhysRevLett.120.050507,doi:10.1126/science.aau0818,Guo2020,PhysRevResearch.3.033043,Lonard2023,Yao2023}, the stability of such phases in the thermodynamic limit remains under debate~\cite{PhysRevB.95.155129,PhysRevLett.124.186601,PhysRevE.102.062144,Abanin2021,PhysRevB.103.024203,PhysRevLett.127.230603,PhysRevB.105.224203,10.21468/SciPostPhys.12.6.201,PhysRevB.105.174205,PhysRevX.13.011041,PhysRevLett.131.106301,PhysRevB.110.014205,PhysRevLett.133.126502,PhysRevB.109.L081117} (for recent reviews of MBL, see \cite{RevModPhys.91.021001,Sierant2025}).

Beyond strong ETH violations, \emph{quantum many-body scars} (QMBS) have recently emerged as a distinct paradigm of `weak' ETH violation where a few nonthermal eigenstates are embedded in an otherwise thermal spectrum~\cite{Serbyn2021, Papic2022, moudgalya2022quantum, annurev:/content/journals/10.1146/annurev-conmatphys-031620-101617}. The key signature of QMBS is long-lived revivals following quenches from special initial states, which have been observed in diverse experimental platforms including Rydberg atom arrays~\cite{Bernien2017,doi:10.1126/science.abg2530,Zhao2025HSF}, ultracold atoms in optical lattices~\cite{PhysRevResearch.5.023010}, and superconducting circuits~\cite{Zhang2023,HangDong2023}. Theoretical works have identified a multitude of QMBS mechanisms, including the existence of non-thermal eigenstates with special algebraic properties~\cite{moudgalya2018exact, PhysRevB.98.235156, Turner2018, Turner2018PRB, PhysRevLett.122.220603, Surace2020, PhysRevLett.123.147201}, the semiclassical nature of QMBS dynamics~\cite{Ho2019,Michailidis2020,Turner2021,Pizzi2025}, projector embeddings~\cite{PhysRevLett.119.030601,Lin2019,McClarty2020,Lee2020,ivanov2025exactarealawscareigenstates}, and schemes that unify some of those~\cite{PhysRevB.101.195131,Pakrouski2020,PhysRevResearch.2.043305, PhysRevLett.126.120604,Gotta2023,Moudgalya2024}. Recently, significant progress has also been made on general-purpose diagnostics of QMBS~\cite{PhysRevLett.131.020402,Feng2025,glc5-hv2m,ren2025scarfinderdetectoroptimalscar,petrova2025findingperiodicorbitsprojected} and extending QMBS from isolated to open quantum systems \cite{Bua2019,PhysRevX.13.031013,10.21468/SciPostPhys.15.2.052,PhysRevLett.132.150401,PhysRevB.109.054311,PhysRevLett.133.216601,zn9v-k73w,ma2025liouvilleanspectraltransitionnoisy}. These advances in QMBS have revealed many ways of ``shielding'' nonthermal states from a thermal background. 

By contrast, in this paper we uncover a converse phenomenon: the emergence of thermal states within a predominantly nonthermal spectrum. Motivated by the general picture of QMBS, where nonthermal eigenstates reside in an invariant subspace defined as the null space of a certain Hamiltonian term, we introduce a complementary framework described by the Hamiltonian $H = H' + H_{\text{dis}}$, where the disorder term $ H_{\text{dis}}$ penalizes specific local configurations. This produces a low-energy manifold free of such configurations. The clean term $H'$ hybridizes states within this manifold and couples them to its complement, thereby generating a small set of highly entangled states which are largely supported within the low-energy manifold. We demonstrate their thermal character through level statistics, entanglement growth after a quantum quench, and dynamics of local observables. As the disorder strength increases, the bulk of the spectrum evolves from being thermal to nonthermal, while the highly entangled thermal states emerge within a narrow band near the spectrum center. Importantly, these thermal states are \emph{induced} by the disorder term and are absent in the clean limit, and we quantitatively relate them to the null space of $H_{\text{dis}}$. We verified the above framework on a variety of spin-1 models, including Heisenberg, XY, and Affleck-Kennedy-Lieb-Tasaki (AKLT) models with several types of disorder; the disordered AKLT model is analyzed in detail below, while additional results for the other models are presented in the Supplementary Material~\cite{SM}. 

\emph{Model.---}We illustrate the above scenario with the disordered AKLT model given by the Hamiltonian
\begin{equation}
  H = H_{\text{AKLT}} + H_{\text{dis}}, \label{Model_AKLT}
\end{equation}
where the standard AKLT Hamiltonian takes the form \cite{PhysRevLett.59.799}
\begin{equation}
  H_{\text{AKLT}} = \sum_{j = 1}^{L - 1} \left( \frac{1}{3} + \frac{1}{2}\vec{S}_{j} \cdot \vec{S}_{j + 1} + \frac{1}{6}(\vec{S}_{j} \cdot \vec{S}_{j + 1})^2 \right).
\end{equation}
Here, $\vec{S}_j$ is a vector of spin-1 operators on site $j$, with the local basis states denoted by $|1\rangle$, $|0\rangle$ and $|\bar{1}\rangle$. We adopt open boundary conditions and assume an even system size $L$. The disorder term $H_{\text{dis}}$ incorporates a three-body interaction
\begin{equation}\label{eq:Hdis}
  H_{\text{dis}} = \sum_{j = 2}^{L - 1} \Delta_j S_{j - 1}^{z} S_{j}^{z} S_{j + 1}^z \left(1 + V_0 S_{j - 1}^z + V_1 S_{j}^z + V_2 S_{j + 1}^z \right),
\end{equation}
where $\Delta_j$ are independent and identically distributed random variables uniformly sampled from $ [-W, W] $, with $W$ being the overall disorder strength. The term in the bracket ensures distinct disorder values for configurations such as $ |11\bar{1}\rangle_{j-1, j, j+1} $, $ |1\bar{1}1\rangle_{j-1, j, j+1} $, and $ |\bar{1}11\rangle_{j-1, j, j+1} $, so that the disorder acts broadly in the Hilbert space. We fix $V_0 = -0.05 $, $ V_1 = 0 $, and $ V_2 = 0.15 $ in all simulations. By varying $W$, we analyze the spectral properties of the model using exact diagonalization. For each system size, we perform $10$--$200$ disorder realizations for a given disorder strength, denoted as $[\cdot]$ in subsequent figures and analysis.

A few remarks are in order. In constructing the disorder term, we ensure that it exhibits two distinct features: (i) an exponentially large null space that serves as the low-energy manifold and hosts potential thermal states; and (ii) outside this null space, the states are either weakly degenerate or non-degenerate, making the background states more likely to exhibit low entanglement \cite{SM}. While we chose the three-body disorder term in Eq.~(\ref{eq:Hdis}) as a representative case, we note that similar behavior can arise from two-body disorder terms constructed following the aforementioned recipe~\cite{SM}. Furthermore, while the AKLT Hamiltonian $ H_{\text{AKLT}} $ preserves both SU(2) symmetry and bond-centered inversion symmetry \cite{moudgalya2018exact}, the disorder term $ H_{\text{dis}} $ explicitly breaks both symmetries. This leaves only the conservation of the total spin projection along the $ z $-axis, denoted as $ S_{\text{tot}}^z $. Unless specified otherwise, we restrict our analysis to the largest symmetry sector, corresponding to $ S_{\text{tot}}^z = 0 $. We also note that, since $H_{\text{dis}}$ annihilates any local spin configurations with $|0\rangle$ in three consecutive spins, the previously found QMBS eigenstates of $ H_{\text{AKLT}}$~\cite{moudgalya2018exact,PhysRevB.108.195133} are no longer eigenstates of $H$. 

\begin{figure}[t]
  \begin{center}
  \includegraphics[width=0.48\textwidth]{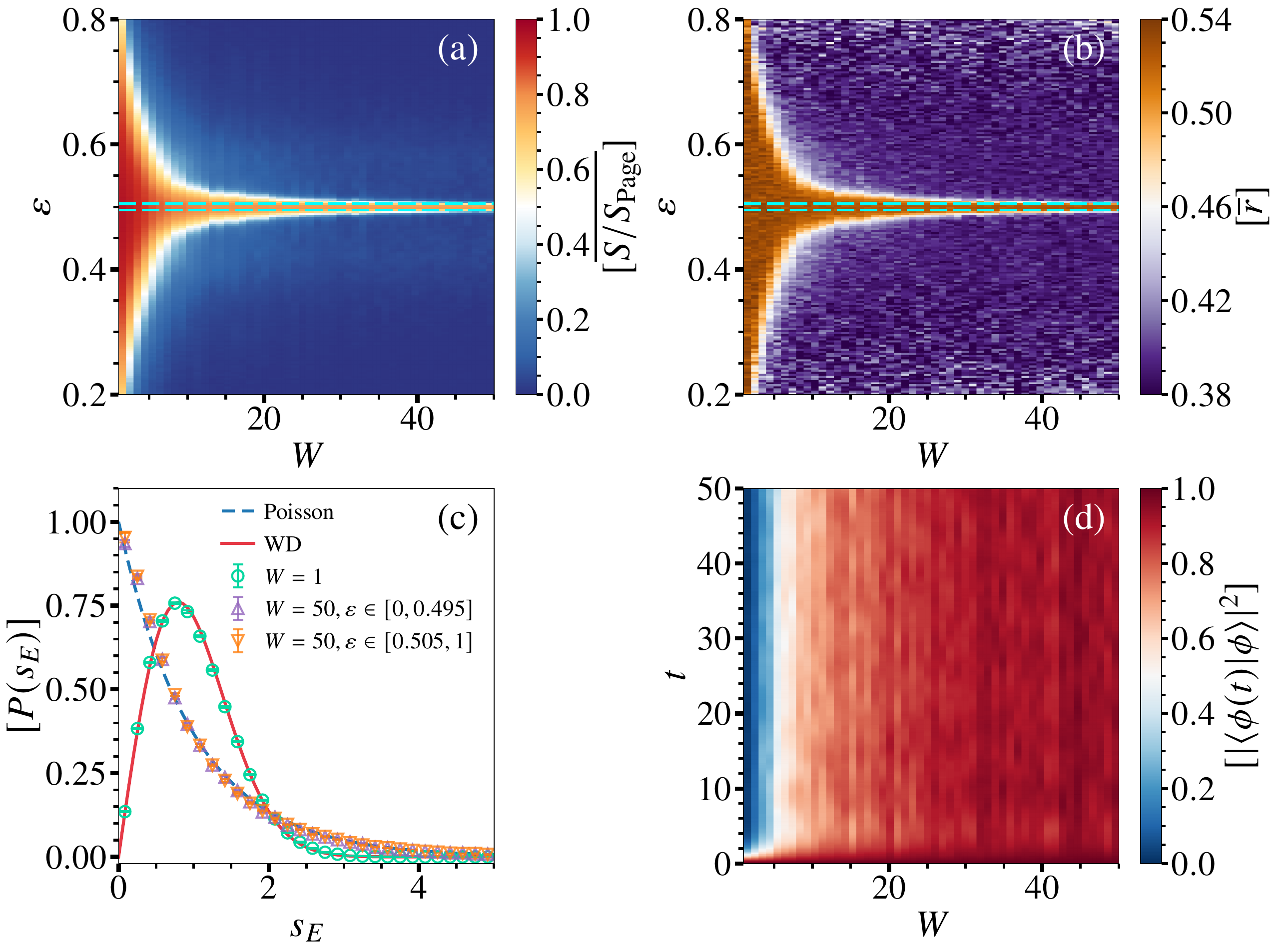}
  \end{center}
  \par
  \renewcommand{\figurename}{Fig.}
\caption{Thermal to nonthermal crossover in the disordered AKLT model [Eq.~(\ref{Model_AKLT})] at system size $L=10$ as a function of disorder strength $W$. Panels (a-b) are the energy-resolved, disorder-averaged EE (a) and level spacing ratio (b). The two dashed cyan lines represent $\varepsilon = 0.495$ and $\varepsilon = 0.505$. (c) Disorder-averaged level statistics at typical disorder strengths. The green dots correspond to the level spacing ratio $[\bar{r}] \approx 0.531$. The purple (orange) triangles indicate $[\bar{r}] \approx 0.400$ ($[\bar{r}] \approx 0.399$). The blue dashed line and red solid line stand for the Poisson and GOE distribution. We consider the central $80\%$ of states across the whole spectrum. (d) Fidelity dynamics of the state $|\phi\rangle = |\bar1 1\bar1\bar111\bar1\bar111\rangle$ for various disorder strengths.}
  \label{Fig_thermal_nonthermal}
  \end{figure}

\emph{Thermal to nonthermal crossover of the background.---} 
We first analyze the entanglement entropy (EE) of the eigenstate of the model in Eq.~(\ref{Model_AKLT}) to assess thermalization properties at different disorder strengths. The bipartite EE is given by 
\begin{equation}
  S = -\operatorname{Tr}_{\mathcal{A}}(\rho_{\mathcal{A}} \log\rho_{\mathcal{A}}),
\end{equation}
where \( \rho_{\mathcal{A}} = \operatorname{Tr}_{\mathcal{B}} \rho \) is the reduced density matrix of subsystem \(\mathcal{A}\) after tracing out the rest of the system \(\mathcal{B}\) (we assume $\mathcal{A}$ and $\mathcal{B}$ are of equal size). For states residing around the middle of the spectrum, their EE approaches the Page value $S_{\text{Page}} \simeq \ln(D_{\mathcal{A}}) - 0.5D_{\mathcal{A}}/D_{\mathcal{B}}$, where $D_{\mathcal{A}}$ and $D_{\mathcal{B}}$ stand for the Hilbert space dimension of the respective subsystems \cite{PhysRevLett.71.1291}. To compare EE across disorder realizations, we introduce the energy density of $E_n$, 
$
  \varepsilon_n = ({E_n - {E_{\text{min}}^\prime}})/({E_{\text{max}} - {E_{\text{min}}^\prime}}),
$
where \( E_n \) denotes the \( n \)'th eigenvalue in ascending order, and \( E_{\text{max}} \) is the maximum energy of the spectrum. Since the minimum and maximum energies of the spectrum vary significantly over different disorder realizations, we choose \(E_{\min}^\prime\equiv -E_{\max}+L\) so as to pin states with \(E_n\approx L/2\) to \(\varepsilon_n\approx0.5\), ensuring a consistent reference frame. For clearer illustration of the energy dependence of EE, we also perform an average over eigenstates within small energy windows across the spectrum, denoted by $\overline{\cdot}$. Figure \ref{Fig_thermal_nonthermal}(a) shows disorder- and state-averaged EE with varying disorder strengths. At weak disorder, the bulk spectrum remains highly entangled, similar to the clean limit at $\Delta=0$, indicating thermalization. As the disorder strength increases, we observe an evolution from high to low entanglement for a majority of states. Notably, high entanglement states simultaneously emerge in a narrow spectral window near the spectrum center. 

We further calculate the energy-resolved level statistics \cite{Guo2020,PhysRevB.91.081103} to illustrate that the crossover from high- to low-entanglement backgrounds corresponds to a switch from thermal to non-thermal behavior as the disorder increases. The level spacing ratio is defined as $r = \langle \min(s_n, s_{n-1}) /  \max(s_n, s_{n-1}) \rangle$, with \( s_n = E_{n+1} - E_n \) denoting the nearest-neighbor energy spacing and \( \langle \cdot \rangle \) representing the spectral average. As shown in Fig.~\ref{Fig_thermal_nonthermal} (b), the disorder-averaged level spacing ratio $[\bar{r}]$ of the background decreases across the energy spectrum as the disorder strength increases, mirroring the behavior of EE [see Fig.~\ref{Fig_thermal_nonthermal}(a)]. For weak disorder, the $[\bar{r}]$ remains high throughout the spectrum, approaching the Gaussian orthogonal ensemble (GOE) value of $r_{\text{GOE}} \approx 0.536$ \cite{PhysRevLett.110.084101}, indicating thermalization. Conversely, as the disorder strength increases, a large portion of the spectrum undergoes a crossover to low $[\bar{r}]$, characteristic of the Poisson value  $r_{\text{Poisson}} \approx 0.386$, suggesting non-ergodic behavior. This evolution is further elucidated in Fig.~\ref{Fig_thermal_nonthermal}(c), where we show the disorder-averaged level statistics computed over energy windows representing the background states. At weak disorder, the level statistics closely follow the GOE distribution, with corresponding level spacing ratio close to $r_{\text{GOE}}$, confirming quantum chaotic behavior \cite{D'Alessio03052016}. Under strong disorder, the statistics of the background states in both selected energy groups resemble a Poisson distribution with their level spacing ratio converging to $r_{\text{Poisson}}$, indicating non-thermal behavior.

Finally, we examine the fidelity dynamics of a specific initial state, $|\phi \rangle = |\bar1 1\bar1\bar111\bar1\bar111\rangle$, which approximates an infinite-temperature state \cite{moudgalya2022quantum} and has the disorder term acting on each site. As depicted in Fig.~\ref{Fig_thermal_nonthermal}(d), the fidelity rapidly decays to a vanishingly small value under weak disorder, indicating a thermalizing background. In contrast, the fidelity remains high at long times under strong disorder, signaling non-ergodic behavior for the background states. These fidelity results, combined with observed crossover in entanglement entropy and level statistics, further support the conclusion that the background states evolve from thermal to nonthermal as disorder strength increases.

\begin{figure}[!t]
  \begin{center}
  \includegraphics[width=0.48\textwidth]{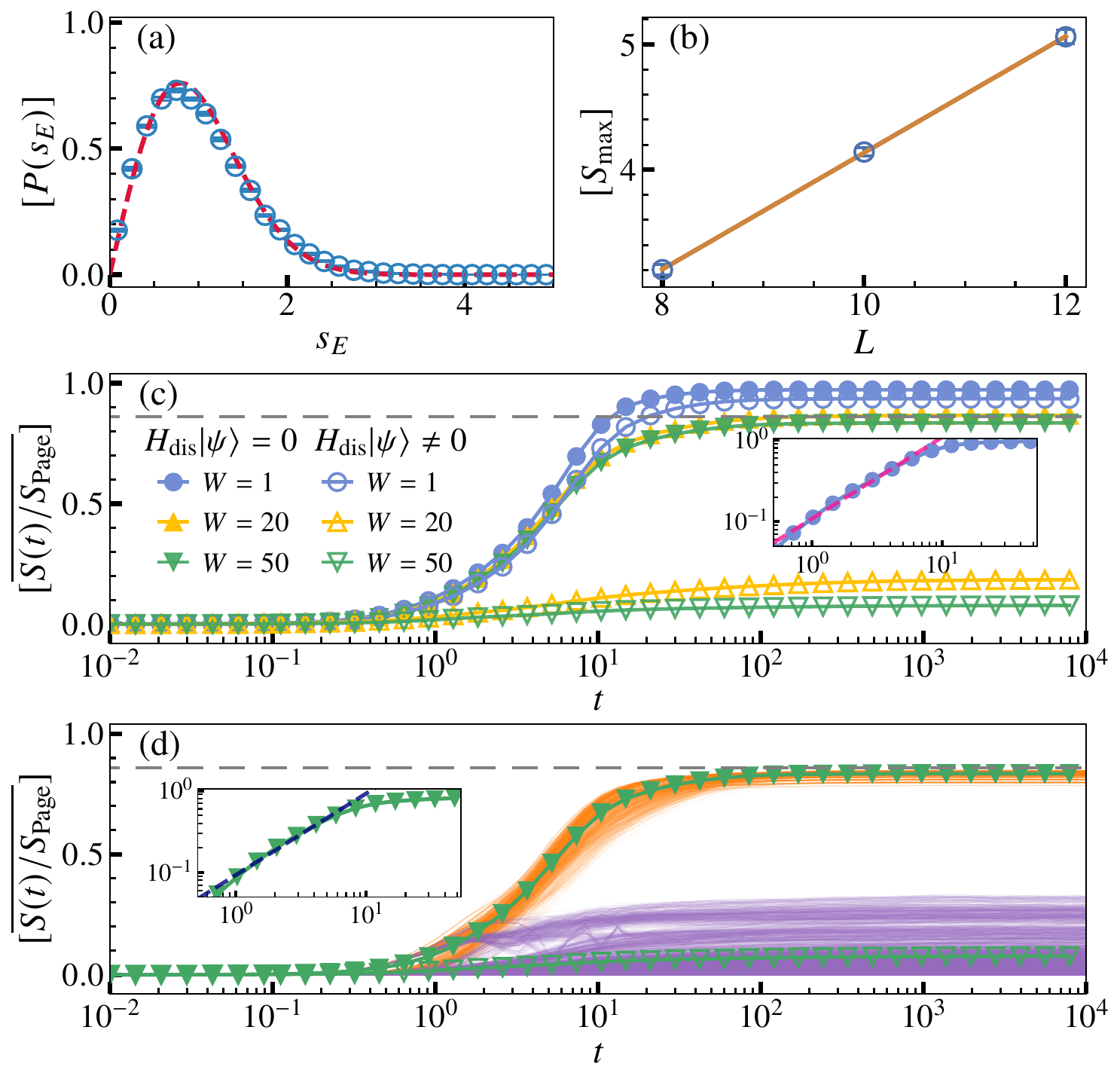}
  \end{center}

  \par
  \renewcommand{\figurename}{Fig.}
  \caption{Thermal properties of the emerged highly entangled states. (a) Disorder-averaged level statistics under $L=10$, $W=50$ and $\varepsilon \in [0.495, 0.505]$, with corresponding $[\bar{r}] \approx 0.523$. (b) Scaling of the disorder-averaged maximum EE $[S_{\text{max}}]$ under $W=50$ over system size $L$. The yellow line gives the linear fitting result. (c) State- and disorder-averaged time evolution of EE initiated from product states, shown for three representative disorder strengths and two distinct groups of initial states. The gray dashed horizontal line indicates the modified saturation value $S^{\text{null}}$. The inset displays a linear fitting for $[\overline{S(t)/S_{\text{Page}}}]$ using initial states inside the null space of $H_{\text{dis}}$ under $W = 1$, with both axes in the logarithm scale. (d) State-resolved disorder-averaged time evolution of EE under $W = 50$. The orange (purple) lines give the state-resolved time evolution of EE for states inside (outside) the null space of $H_{\text{dis}}$. The green dotted lines are reproduced from (c) for reference. The gray dashed line indicates the value of $S^{\text{null}}$. The inset presents the linear fitting result of the initial growth of $[\overline{S(t)/S_{\text{Page}}}]$ for states inside the null space.}
  \label{Fig_highEE}
\end{figure}

\emph{The emerging thermal states on top of low-entangled spectrum.---}
We have demonstrated that states within the energy shell $\varepsilon \in [0.495, 0.505]$ exhibit atypically high EE even under strong disorder [see Fig.~\ref{Fig_thermal_nonthermal}(a)], accompanied by level-spacing ratios approaching $r_{\text{GOE}}$ [see Fig.~\ref{Fig_thermal_nonthermal}(b)], both suggesting potential thermal nature. In the following, we substantiate the emergence of the thermal states on top of the non-thermal sea using level statistics and the time evolution of both EE and site-resolved magnetization following a quantum quench.

We first compute the level statistics of eigenstates within the energy interval $\varepsilon \in [0.495, 0.505]$. As shown in Fig.~\ref{Fig_highEE}(a), these states exhibit a Wigner-Dyson (WD) distribution, in stark contrast to states outside this window. The corresponding level spacing ratio for these states approaches the GOE value, consistent with their chaotic nature. Additionally, the disorder-averaged maximum EE $[S_{\text{max}}]$ grows linearly with system size $L$, as depicted by Fig.~\ref{Fig_highEE}(b). This volume-law scaling behavior confirms the thermal nature of these states. 

To further confirm the thermal nature of the emergent states, we analyze their quench dynamics by examining two key observables: the time evolution of the disorder- and state-averaged EE, $[\overline{S(t)/S_{\text{Page}}}]$, and of the disorder-averaged, site-resolved magnetization, $[\langle S_i^z(t)\rangle]$. We choose the product states as initial states and categorize them into two distinct groups: those lying within the null space of $H_{\text{dis}}$ (i.e. $H_{\text{dis}} |\psi\rangle = 0$) and those outside it (i.e. $H_{\text{dis}} |\psi\rangle \neq 0$). This classification is motivated by our earlier observation that the highly entangled states remain robust across varying disorder strengths, implying an intrinsic link to the null space of $H_{\text{dis}}$. The validity of this partition will be verified in the next section.

Previous studies have established that entanglement growth after a quantum quench in one-dimensional chaotic systems typically exhibits ballistic behavior \cite{PhysRevLett.111.127205,PhysRevX.7.031016}. In our case, we find that the temporal evolution of EE displays distinctly different behaviors depending on both the disorder strength and whether the initial state lies within the null space of $H_{\text{dis}}$. As shown in Fig.~\ref{Fig_highEE}(c), during early time (i.e., $t \ll 1$), both groups under all disorder strengths display similar initial entanglement growth, with entanglement primarily built between the two spins adjacent to the central bond \cite{PhysRevLett.111.127205}. Under weak disorder ($W=1$), the average EE of both groups grows linearly [see the inset of Fig.~\ref{Fig_highEE}(c)] and saturates near $S_{\text{Page}}$, indicating thermalization \cite{PhysRevLett.111.127205,PhysRevX.7.031016}. In contrast, for larger disorder strength in the long time limit, a significant discrepancy between the two groups emerges. As further presented in Fig.\ref{Fig_highEE}(d), all states outside the null space display slow EE growth that saturates well below $S_{\text{Page}}$, signaling non-ergodic nature of major eigenstates. Meanwhile, while states inside the null space still exhibit initial ballistic growth [see Fig.~\ref{Fig_highEE}(d) and Supplementary Materials \cite{SM}], implying chaotic behavior, their late-time EE saturates at a value slightly lower than $S_{\text{Page}}$, suggesting exploration of a restricted Hilbert space subspace. To capture this behavior, we introduce a modified saturation value $S^{\text{null}}$, where the corresponding $D_{\mathcal{A}}$ and $D_{\mathcal{B}}$ are computed within the null space, following similar procedures as in Refs. \cite{PhysRevLett.124.207602,PhysRevB.109.014212}. As illustrated in Fig.~\ref{Fig_highEE}(c) and (d), $S^{\text{null}}$ provides an accurate estimation for the late-time entanglement of states within the null space, indicative of a complete exploration of the null space.

\begin{figure}[!t]
  \begin{center}
  \includegraphics[width=0.48\textwidth]{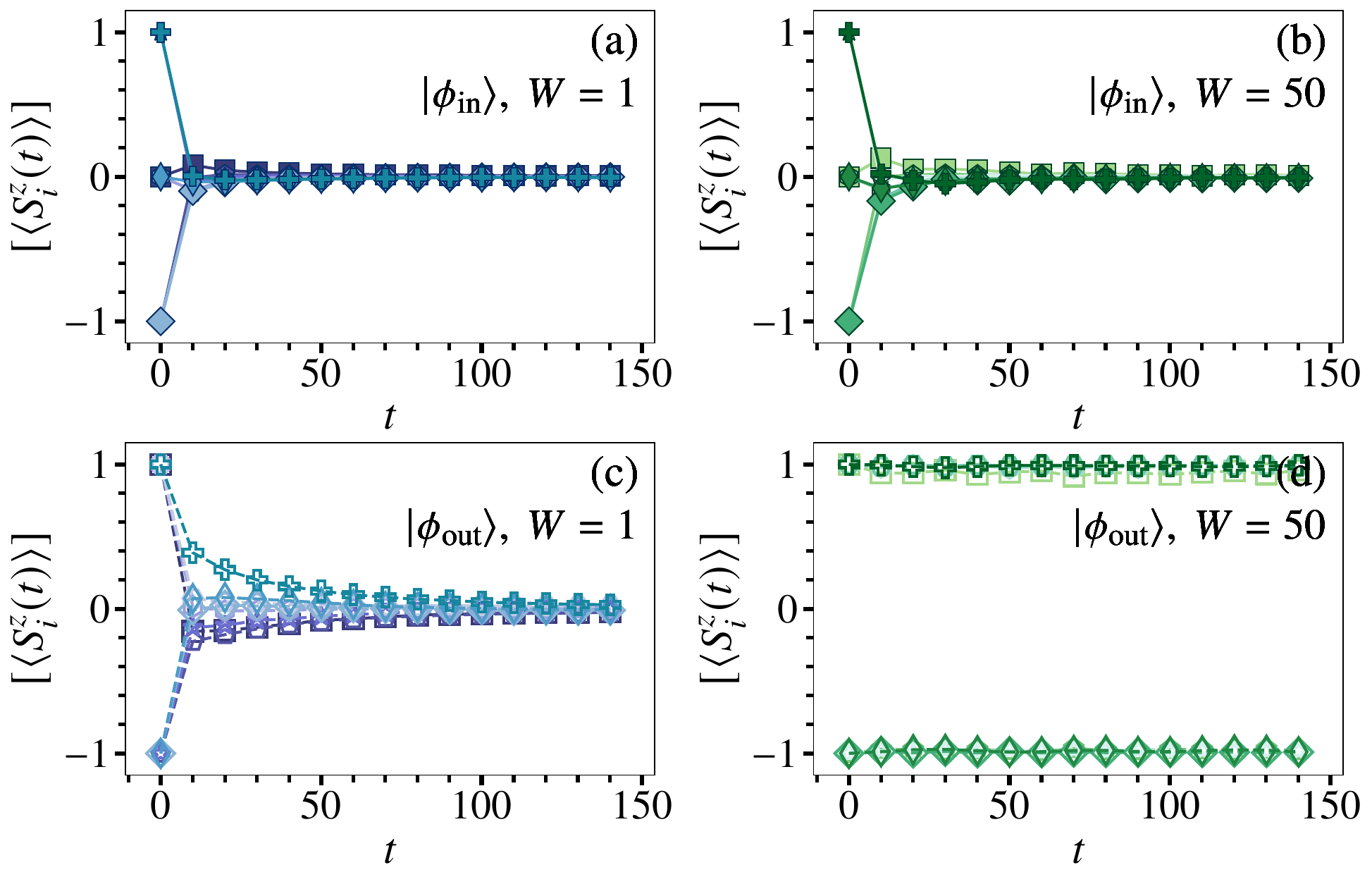}
  \end{center}
  \par
  \renewcommand{\figurename}{Fig.}
  \caption{Disorder-averaged time evolution of site-resolved magnetization. (a) and (b) are calculated for the initial state $|\phi_{\text{in}}\rangle$ at representative disorder strength $W$, while (c) and (d) correspond to the initial state $|\phi_{\text{out}}\rangle$. Lines with varying color darkness and markers track the evolution of individual sites. Note that we only display the central $8$ spins for $L=10$.}
  \label{Fig_M_t}
\end{figure}

Beyond evolution of EE, time evolution of disorder-averaged site-resolved magnetization $[\langle S_i^z(t)\rangle]$ can also capture thermal behavior, as demonstrated experimentally \cite{smith2016many}. In a thermalizing system, a quantum quench will erase memory regarding all original information of a state. Yet in non-ergodic systems, observables do not relax to thermal values and information can be preserved over time \cite{smith2016many,PhysRevB.82.174411}. Akin to the setup in \cite{smith2016many}, we consider two initial states $|\phi_{\text{in}}\rangle = |10\bar110\bar1\bar1010\rangle$ and $|\phi_{\text{out}}\rangle = |\bar1 1\bar1\bar111\bar1\bar111\rangle$, both residing close to the center of the spectrum without disorder. While $|\phi_{\text{in}}\rangle$ is unaffected by $H_{\text{dis}}$ and thus signifying states within the null space of $H_{\text{dis}}$, $|\phi_{\text{out}}\rangle$ is impacted on every site and corresponds to states outside of the null space. Under weak disorder, both states have their $[\langle S_i^z(t)\rangle]$ evolving to thermal equilibrium values, losing memory of their initial spin pattern [see Fig.~\ref{Fig_M_t} (a), (c)]. As we increase the disorder strength, site-resolved magnetization of $|\phi_{\text{in}} (t)\rangle$ persists to thermalize around $[\langle S_i^z (t)\rangle] \approx 0$ [see Fig.~\ref{Fig_M_t} (b)]. In contrast, as shown in Fig.~\ref{Fig_M_t} (d), for $|\phi_{\text{out}} (t)\rangle$, the spin configuration is nearly perfectly preserved under strong disorder. These observations confirm the emergence of highly entangled thermal states embedded in the otherwise non-thermal spectrum.

\emph{The origin of disorder-induced thermal states.---}Within the framework $H=H'+H_{\text{dis}}$, we now investigate how the thermal states arise and where they live in Hilbert space. We first test whether these states are already present in the clean limit ($W=0$). By calculating the overlap between the maximally entangled thermal state acquired under large disorder strength ($W=50$) and the eigenstates of the model Hamiltonian in the clean limit ($W=0$), we find that the overlap is vanishingly small for all clean limit eigenstates [see the inset of Fig.~\ref{Fig_Overlap}]. This demonstrates that the thermal states are not inherited from the clean limit but are instead induced by disorder.

To assess the extent to which these thermal states reside in the null space of $H_{\text{dis}}$, we project a given eigenstates $|\psi \rangle$ onto the orthonormal basis $\{ |\psi_i^{\text{null}} \rangle \}$ spanning the null space of $H_{\text{dis}}$, and examine the total fidelity $\sum_{i} |\langle \psi | \psi_i^{\text{null}} \rangle|^2$. As shown in Fig.~\ref{Fig_Overlap}, within the narrow energy window $\varepsilon \in [0.495, 0.505]$, most states exhibit both high fidelity and large EE, whereas both quantities become vanishingly small outside this window. These results indicate that the highly entangled thermal states are primarily composed of superpositions of disorder-free configurations within the null space of $H_{\text{dis}}$. However, since the maximal fidelity in Fig.~\ref{Fig_Overlap} remains slightly below unity, we conclude that these thermal states are not entirely confined to the null space but include small admixtures from its complement, i.e., $H_{\text{dis}}|\psi\rangle \neq 0$. This property highlights the difference between the thermal states discussed here and the previously studied ``disorder-annihilated'' states which obey $H_{\text{dis}}|\psi\rangle=0$ \cite{PhysRevLett.125.240401, PhysRevB.106.214201, PhysRevB.108.L100202, PhysRevA.109.023310}. The same behavior holds across disorder realizations \cite{SM}.

\begin{figure}[!t]
  \begin{center}
  \includegraphics[width=0.48\textwidth]{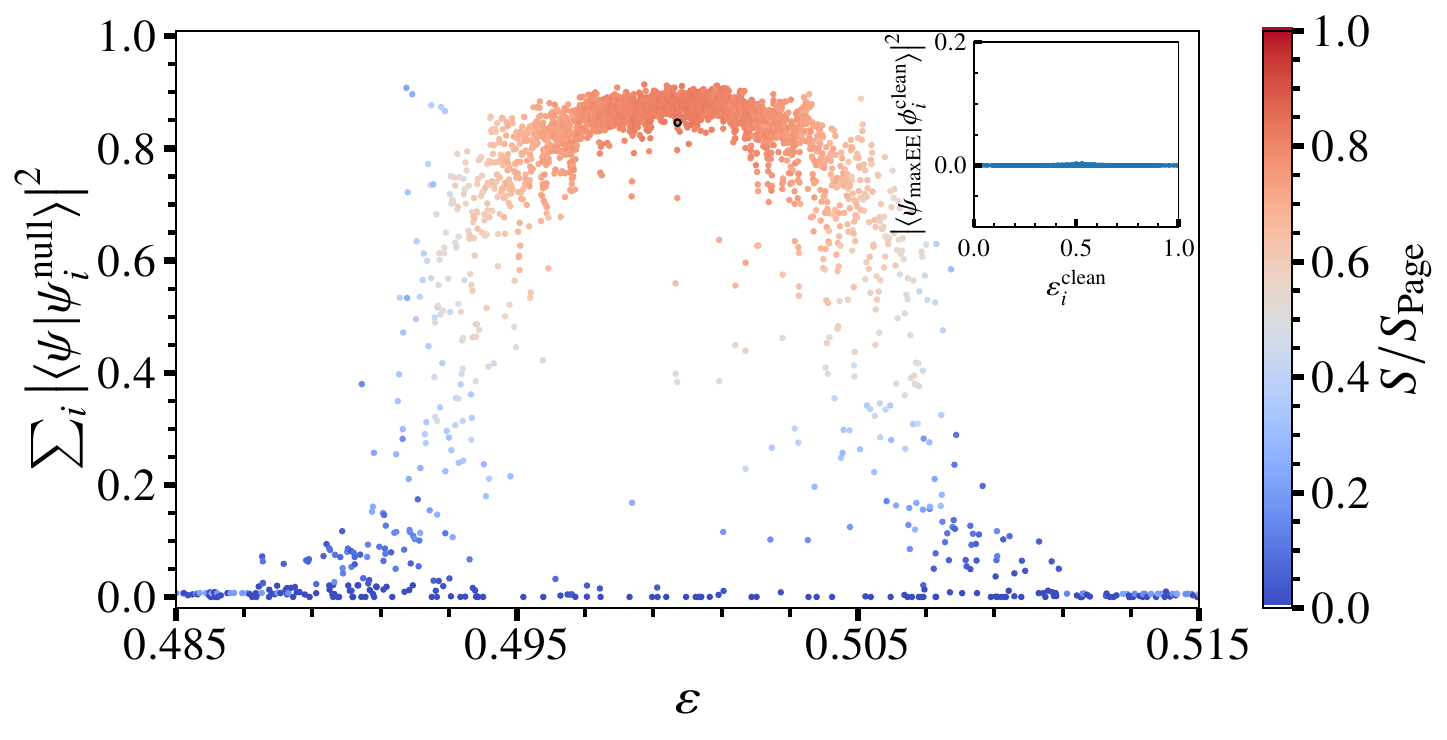}
  \end{center}
  \par
  \renewcommand{\figurename}{Fig.}
  \caption{Total fidelity between each eigenstate $|\psi\rangle$ of the full Hamiltonian and the orthonormal basis ${ |\psi_i^{\text{null}} \rangle }$ spanning the null space of the disorder term $H_{\text{dis}}$, calculated for a representative disorder realization at $W = 50$. The black circle indicates the maximally entangled state $|\psi_{\text{maxEE}}\rangle$. 
  Inset shows overlaps between $|\psi_{\text{maxEE}}\rangle$ and eigenstates $\{|\phi_i^{\text{clean}}\rangle\}$ of the model Hamiltonian under clean limit, with $\varepsilon_i^{\text{clean}}$ being the energy density of $|\phi_i^{\text{clean}}\rangle$.}
  \label{Fig_Overlap}
  \end{figure}

\emph{Conclusions and discussion.---}In this work, we proposed a generic framework for the emergence of a narrow band of thermal eigenstates embedded within the otherwise low-entanglement spectrum in a broad class of disordered spin-$1$ systems. The thermal character of the emergent highly-entangled states is confirmed by the chaotic level statistics, time evolution of the EE and site-resolved magnetization in quenches from product states within the null space of the disorder term. Furthermore, using the total fidelity as an indicator, we illustrated that the highly entangled thermal states are predominantly supported by the null space of $H_{\text{dis}}$.

Given recent advances in constructing interacting multi-qubit systems using state-of-the-art superconducting quantum simulation platforms \cite{PhysRevLett.129.220501, HangDong2023}, as well as progress in simulating spin-$1$ Hamiltonians with Rydberg atoms \cite{qiao2025} (for theoretical proposals, see \cite{PRXQuantum.6.020332,kunimi2025}), our results above could be readily tested in some of these platforms. 

The framework presented here can be applied to a broad class of spin models, and it generalizes the previous spin-$1/2$ special cases \cite{PhysRevB.109.014212}. In all of these examples, the highly-entangled states are disorder-induced, absent from the clean-limit spectrum, and robustly thermal, thus they are intrinsically random. This sharply contrasts them with the disorder-annihilated states studied in Refs.~\cite{PhysRevLett.125.240401, PhysRevB.106.214201, PhysRevB.108.L100202, PhysRevA.109.023310}. The latter also exhibit relatively high entanglement compared to their background under large disorder, since they are exact, nonthermal eigenstates satisfying $H_{\text{dis}}|\psi\rangle=0$ and pre-existing in the clean limit. However, these disorder-annihilated states are distinct from the thermal states discussed in our paper. 

Finally, our results not only broaden the landscape of ETH breaking systems, but also offer new insight into the breakdown of MBL, even under strong disorder and within finite-size systems. For example, the sharp contrast between the emerging highly-entangled thermal states and the surrounding non-thermal states is reminiscent of a mobility edge---a specific energy density dividing localized and extended states~\cite{10.1063/1.2994815}. Our scenario, however, is distinct from the conventional mobility edge for two reasons: (i) the highly-entangled states coexist with low-entanglement states within the \emph{same} energy window [see Fig.~\ref{Fig_Overlap}]; and (ii) the highly-entangled states appear only within a finite window, rather than being separated from non-thermal states by a single critical energy. Nevertheless, it would be interesting to explore if our construction could be generalized to provide a tractable model for a conventional MBL mobility edge.
 
\begin{acknowledgments}
\emph{Acknowledgment.--- }  We thank helpful discussions with Wen-Da Tang. This work was supported by the National Natural Science Foundation of China (Grant No.92477106), the Fundamental Research Funds for the Central Universities. Z.P. acknowledges support by the Leverhulme Trust Research Leadership Award RL-2019-015 and EPSRC Grant EP/Z533634/1. Statement of compliance with EPSRC policy framework on research data: This publication is theoretical work that does not require supporting research data. This research was supported in part by grant NSF PHY-2309135 to the Kavli Institute for Theoretical Physics (KITP). 
\end{acknowledgments}

\bibliography{ref}
\begin{appendix}

\onecolumngrid
\newpage
\renewcommand{\theequation}{S\arabic{equation}}
\setcounter{equation}{0}
\renewcommand{\thefigure}{S\arabic{figure}}
\setcounter{figure}{0}

\renewcommand{\thesection}{S\arabic{section}}
\renewcommand{\thesubsection}{S\arabic{section}.\arabic{subsection}}

\makeatletter
\renewcommand{\p@subsection}{}  

\newcommand{\appendixsection}[1]{
  \refstepcounter{section}
  \setcounter{subsection}{0}
  \vspace{2ex}
  \noindent\begin{center}
    \textbf{\thesection\quad #1}
  \end{center}
  \vspace{1ex}
}

\newcommand{\appendixsubsection}[1]{
  \refstepcounter{subsection}
  \vspace{1.5ex}
  \noindent\begin{center}
    \textbf{\thesubsection\quad #1}\par
  \end{center}
  \vspace{0.8ex}
}
\makeatother

\begin{center}
  \textbf{\large{Supplementary Materials for\\``{Thermal states emerging from low-entanglement background in disordered spin models}"}}
\end{center}

\maketitle

\appendixsection{Alternative models}

Our observed phenomenon, in which a small fraction of highly entangled states emerge atop a predominantly low-entanglement background, can occur in various spin-$1$ models with diverse clean Hamiltonians and disorder terms. To illustrate this, we present several alternative models that exhibit similar behavior to the one reported in the main text. We first show results obtained with clean Hamiltonians distinct from the AKLT Hamiltonian, and then introduce other disorder terms with the AKLT model.

In Sec.~\ref{sec:DisHei} and Sec.~\ref{sec:DisXY}, we extend our analysis from the AKLT model to other spin-$1$ Hamiltonians, demonstrating the universality of the observed phenomenon in spin-$1$ systems. Specifically, we present calculations in which the AKLT Hamiltonian in Eq.~(\ref{Model_AKLT}) is replaced by the spin-$1$ XY and Heisenberg Hamiltonians.

In Sec.~\ref{sec:Dis1} and Sec.~\ref{sec:Dis2}, we introduce two alternative two-body disorder terms, in addition to the three-body term proposed in the main text. The disorder term in the main text possesses two essential features: (i) it has an exponentially large null space, allowing the majority of spin configurations to thermalize upon adding $H_{\text{AKLT}}$; and (ii) when considered alone, states outside the null space are either non-degenerate or only weakly degenerate, making it more likely that most eigenstates of the total Hamiltonian form a low-entanglement background. The alternative disorder terms we propose share these two essential properties. We apply both to the AKLT model and present the corresponding results.

\appendixsubsection{Disordered Heisenberg model}\label{sec:DisHei}

We first consider the case with a Heisenberg clean Hamiltonian. The total Hamiltonian  now reads
\begin{equation}
    H_{\text{Hei}+\text{dis}} = H_{\text{Hei}} + H_{\text{dis}}, \label{Hei}
\end{equation}
where
\begin{gather}
    H_{\text{Hei}} = \sum_{j=1}^{L-1} S_j^{x}S_{j+1}^x + S_j^{y}S_{j+1}^y + S_j^{z}S_{j+1}^z, \\ H_{\text{dis}} = \sum_{j=2}^{L-1} \omega_j S_{j-1}^z S_j^z S_{j+1}^z(1 + V_0 S_{j-1}^z + V_1 S_{j}^z + V_2 S_{j+1}^z) \label{AKLT_dis}.
\end{gather}
Here, $\omega_j$ are independent and identically distributed random variables uniformly sampled from $[-W, W]$. We set $V_0 = -0.05$, $V_1 = 0$, $V_2 = 0.15$, as in the main text. In order to compare across different disorder strengths, we again introduce the energy density. Unlike the AKLT model, for both the XY model and the Heisenberg model, the medium of the energy spectrum shows around $E \approx 0$, thus we define the energy density $\varepsilon_n$ of the energy $E_n$ as $\varepsilon_n = ({E_n + E_{\text{max}}})/({2E_{\text{max}} })$, with \( E_{\text{max}} \) representing the maximum energy of the spectrum. 

\begin{figure}[htb]
  \begin{center}
  \includegraphics[width=0.95\textwidth]{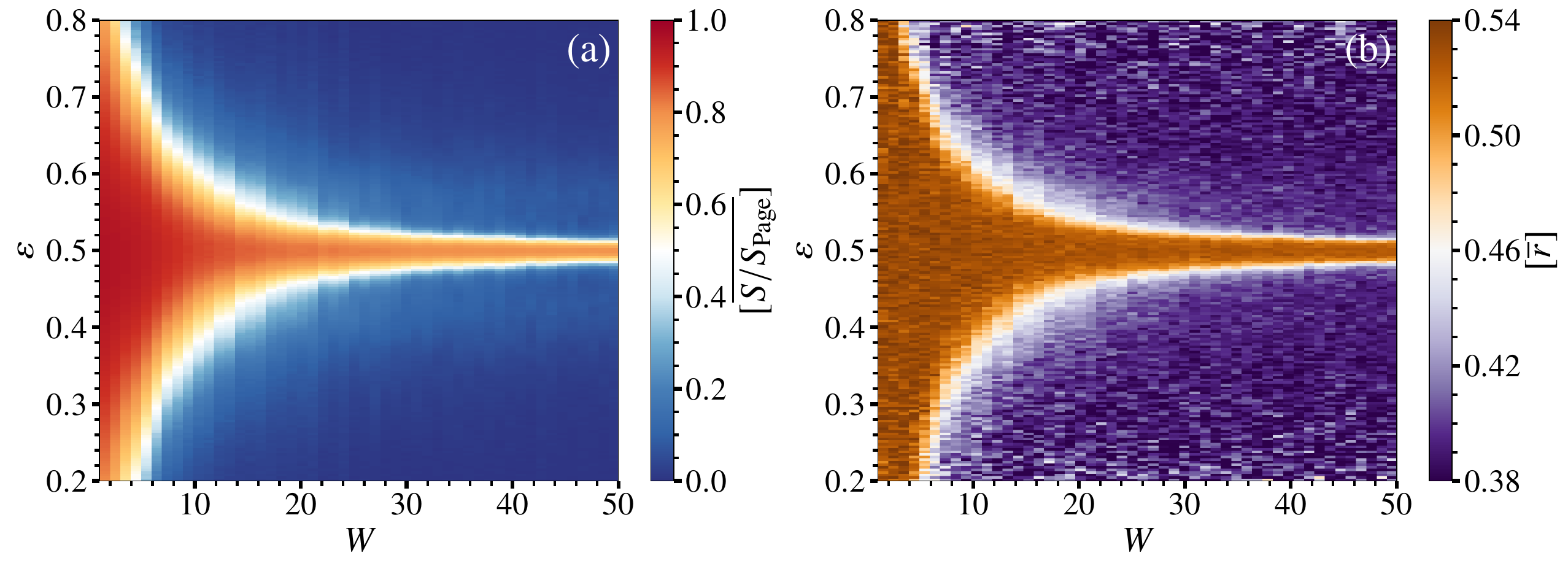}
  \end{center}

  \par
  \renewcommand{\figurename}{Fig.}
  \caption{(a) Energy spectrum and (b) level spacing ratio of the disordered Heisenberg model at $L=10$ and various disorder strengths.}
  \label{Fig_Hei}
\end{figure}

\begin{figure}[htb]
  \begin{center}
  \includegraphics[width=0.95\textwidth]{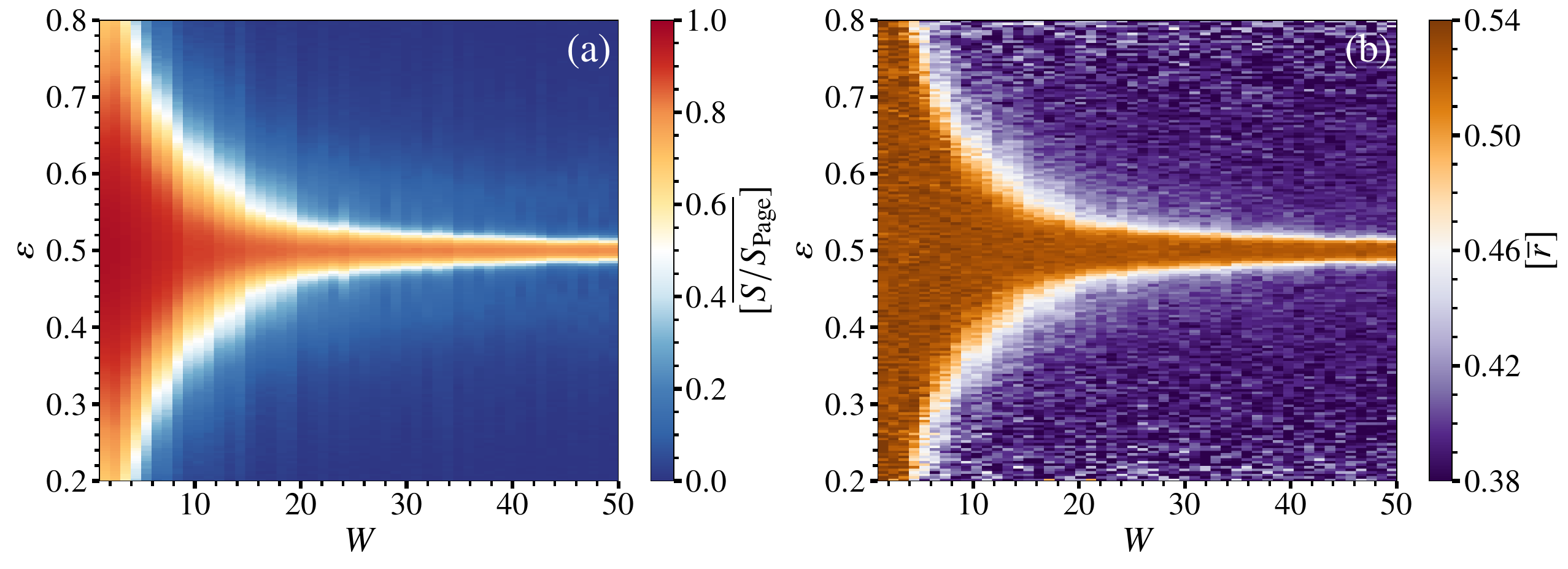}
  \end{center}

  \par
  \renewcommand{\figurename}{Fig.}
  \caption{(a) Energy spectrum and (b) level spacing ratio of the disordered XY model at $L=10$ and various disorder strengths.}
  \label{Fig_XY}
\end{figure}

Here, we present the energy-resolved disorder-averaged energy spectrum and level spacing ratio to illustrate the main features of the system with $L=10$ spins. As shown in Fig.~\ref{Fig_XY}, increasing the disorder strength causes most background states to evolve from high entanglement to low entanglement, accompanied by a shift in the level spacing ratio from the GOE value to the Poisson value, indicative of a thermal-to-nonthermal crossover. Notably, within a narrow energy window near the center of the spectrum, a small fraction of highly entangled states emerge, with the level spacing ratio approaching the GOE value, a hallmark for thermal features. The results for the disordered Heisenberg model are consistent with those obtained for the AKLT model presented in the main text.

\appendixsubsection{Disordered XY model}\label{sec:DisXY}

Here we consider the disordered XY model with the disorder term defined in the main text
\begin{equation}
    H_{\text{XY}+\text{dis}} = H_{\text{XY}} + H_{\text{dis}}, \label{XY}
\end{equation}
where
\begin{equation}
    H_{\text{XY}} = \sum_{j=1}^{L-1} S_j^{x}S_{j+1}^x + S_j^{y}S_{j+1}^y.
\end{equation}
The results of the energy-resolved disorder-averaged energy spectrum and the level spacing ratio of the disordered XY model with $L=10$ are shown in Fig.~\ref{Fig_XY}. The transition from thermal to nonthermal behavior in the background states, as well as the emergence of highly entangled thermal states, are clearly visible---consistent with our results for the disordered AKLT and Heisenberg models. These findings indicate that the existence of a small fraction of highly entangled states on top of a background of lowly entangled states is not unique to the AKLT model. Rather, a similar phenomenon can also be observed in more general spin-$1$ models, such as the XY and Heisenberg models, demonstrating the universality of our findings.

\appendixsubsection{Alternative two-body disorder term: variant  I}\label{sec:Dis1}

The first alternative disorder term we propose takes the form
\begin{equation}
    H_{\text{dis}1} = \sum_{j} \omega_1^{j} \left((S_{j}^{+})^2(S_{j+1}^{-})^2 + (S_{j}^{-})^2(S_{j+1}^{+})^2 \right) + \omega_2^j |1 \circ 1\rangle_{j,j+1,j+2} \langle 1 \circ 1|_{j,j+1,j+2}, \label{dis1}
\end{equation}
where $\omega_1^j$ and $\omega_2^j$ are independent and identically distributed random variables uniformly sampled from $[-W, W]$. Here we use $\circ$ to represent any spin on site $j+1$. The total Hamiltonian reads 
\begin{equation}
    H_{\text{AKLT}+\text{dis1}} = H_{\text{AKLT}} + H_{\text{dis1}}. \label{AKLT_dis1}
\end{equation}
For this model, we still use the energy density defined in the main text $\varepsilon_n = ({E_n - E'_{\text{min}}})/({E_{\text{max}} - E'_{\text{min}}})$, where $E'_{\text{min}} = -E_{\text{max}} + L$ and \( E_{\text{max}} \) represents the maximum energy of the spectrum. 

\begin{figure}[thb]
    \begin{center}
    \includegraphics[width=0.95\textwidth]{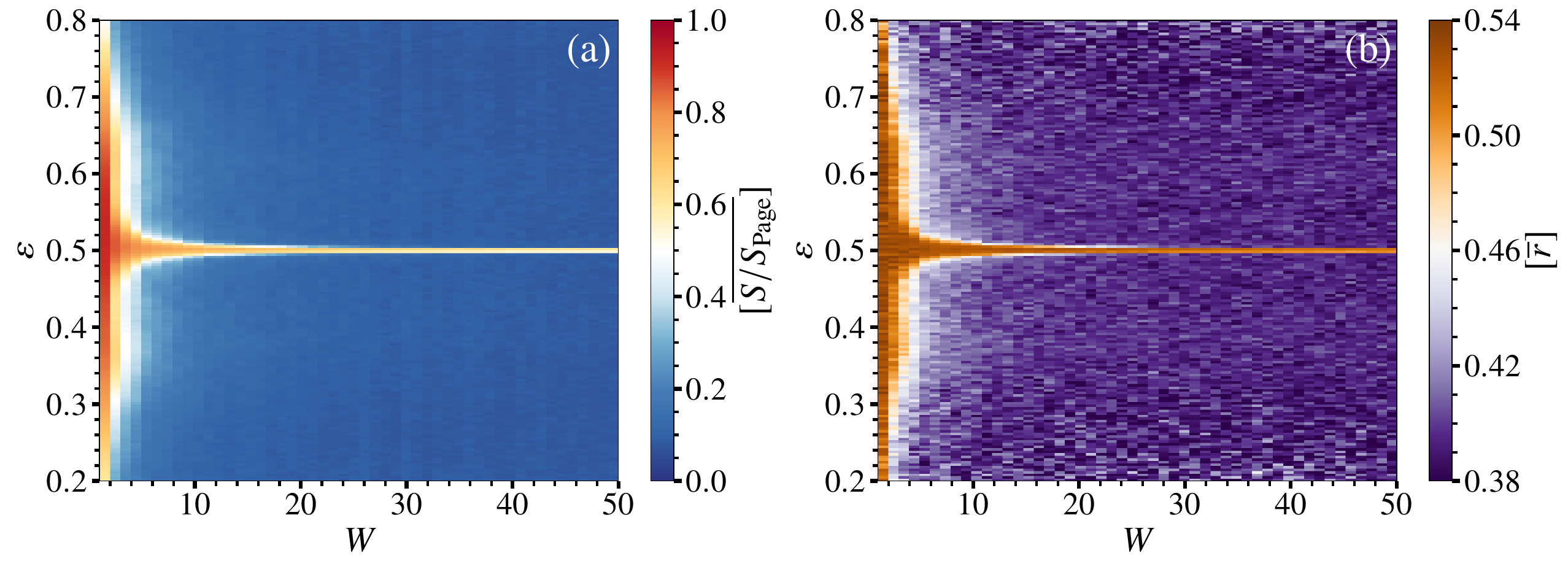}
    \end{center}
    \par
    \renewcommand{\figurename}{Fig.}
    \caption{(a) Energy spectrum and (b) level spacing ratio of the disordered AKLT model in Eq.(\ref{AKLT_dis1}) under various disorder strengths.}
    \label{Fig_AKLT1}
\end{figure}

Energy-resolved disorder-averaged energy spectrum and level spacing ratio calculated based on the disordered AKLT model in Eq.~(\ref{AKLT_dis1}) for $L=10$ are shown in Fig.~\ref{Fig_AKLT1}. As the disorder strength increases, states over a broad energy window shift from high-entanglement to low-entanglement, accompanied by a shift in the level-spacing ratio from the GOE value to the Poisson value, indicating a thermal-to-nonthermal crossover for the background states. In the large disorder regime, a subset of highly entangled states emerges with level-spacing ratios close to the GOE value, signaling the presence of thermal states. These results are consistent with those obtained using the disorder term defined in the main text.

We should note that the high-entanglement part of the spectrum in Fig.~\ref{Fig_AKLT1} appears “thinner” merely because the energy bandwidth in this case is larger than in the aforementioned examples. This is due to
\begin{equation}
    \omega_1^j (S_{j}^{+})^2(S_{j+1}^{-})^2 |\bar{1}1\rangle_{j, j+1} = 4 \omega_1^j |1\bar{1}\rangle_{j, j+1}.
\end{equation}
The disorder strength is effectively quadrupled, leading to a thermal-to-nonthermal transition of the background at relatively small disorder values. For a given disorder strength, the system's energy bandwidth is wider than in comparable cases. While the absolute bandwidth of the highly entangled part is essentially unchanged, its width in energy-density units is reduced due to the larger $E_{\text{max}} - E_{\text{min}}'$. 

\appendixsubsection{Alternative Two-Body Disorder Term: Variant  II}\label{sec:Dis2}
The second alternative disorder term reads
\begin{equation}
    H_{\text{dis}2} = \sum_j \omega_1^j |11\rangle_{j, j+1} \langle11|_{j, j+1} + \omega_2^j |\bar{1}\bar{1}\rangle_{j, j+1} \langle\bar{1}\bar{1}|_{j, j+1} +  \omega_3^j |\bar{1} \circ 1\rangle_{j, j+1, j+2} \langle\bar{1} \circ 1|_{j, j+1, j+2}, \label{dis2}
\end{equation}
where $\omega_1^j$, $\omega_2^j$ and $\omega_3^j$ are independent and identically distributed random variables uniformly sampled from $[-W, W]$. The total Hamiltonian takes the form
\begin{equation}
    H_{\text{AKLT}+\text{dis2}} = H_{\text{AKLT}} + H_{\text{dis2}}. \label{AKLT_dis2}
\end{equation}

\begin{figure}[thb]
    \begin{center}
    \includegraphics[width=0.95\textwidth]{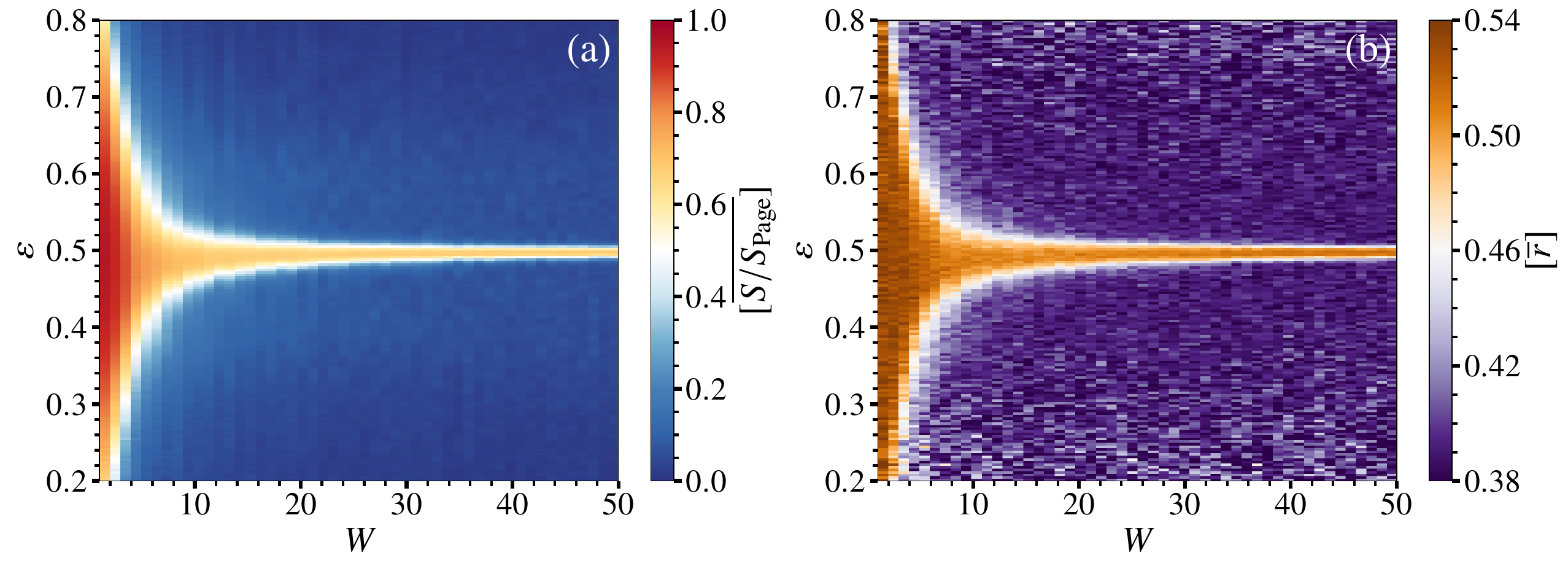}
    \end{center}
    \par
    \renewcommand{\figurename}{Fig.}
    \caption{(a) Energy spectrum and (b) level spacing ratio of the disordered AKLT in Eq.(\ref{AKLT_dis2}) under various disorder strengths.}
    \label{Fig_AKLT2}
\end{figure}

Energy-resolved disorder-averaged energy spectrum and level spacing ratio calculated based on this disordered AKLT model Eq.~(\ref{AKLT_dis2}) under $L=10$ are shown in Fig.~\ref{Fig_AKLT2}. Again, we observe a clear transition of the background states from thermal to nonthermal behavior, along with the emergence of highly entangled thermal states near the center of the spectrum at large disorder strength.

The above results confirm the existence of a broad range of disorder terms capable of generating disorder-induced highly entangled states, provided they meet the two aforementioned criteria. This demonstrates that the phenomenon is robust and pervasive.

\appendixsubsection{Scaling of the null-space dimension}

Here we present the scaling behavior of the null-space dimension for the disorder terms introduced above, reinforcing our claim that all proposed disorder terms possess an exponentially large null space. In what follows, we denote the null-space dimensions of $H_{\text{dis}}$, $H_{\text{dis1}}$, and $H_{\text{dis2}}$ as $D_{\text{dis}}^{\text{null}}$, $D_{\text{dis1}}^{\text{null}}$, and $D_{\text{dis2}}^{\text{null}}$, respectively.

\begin{figure}[!tb]
    \begin{center}
    \includegraphics[width=0.45\textwidth]{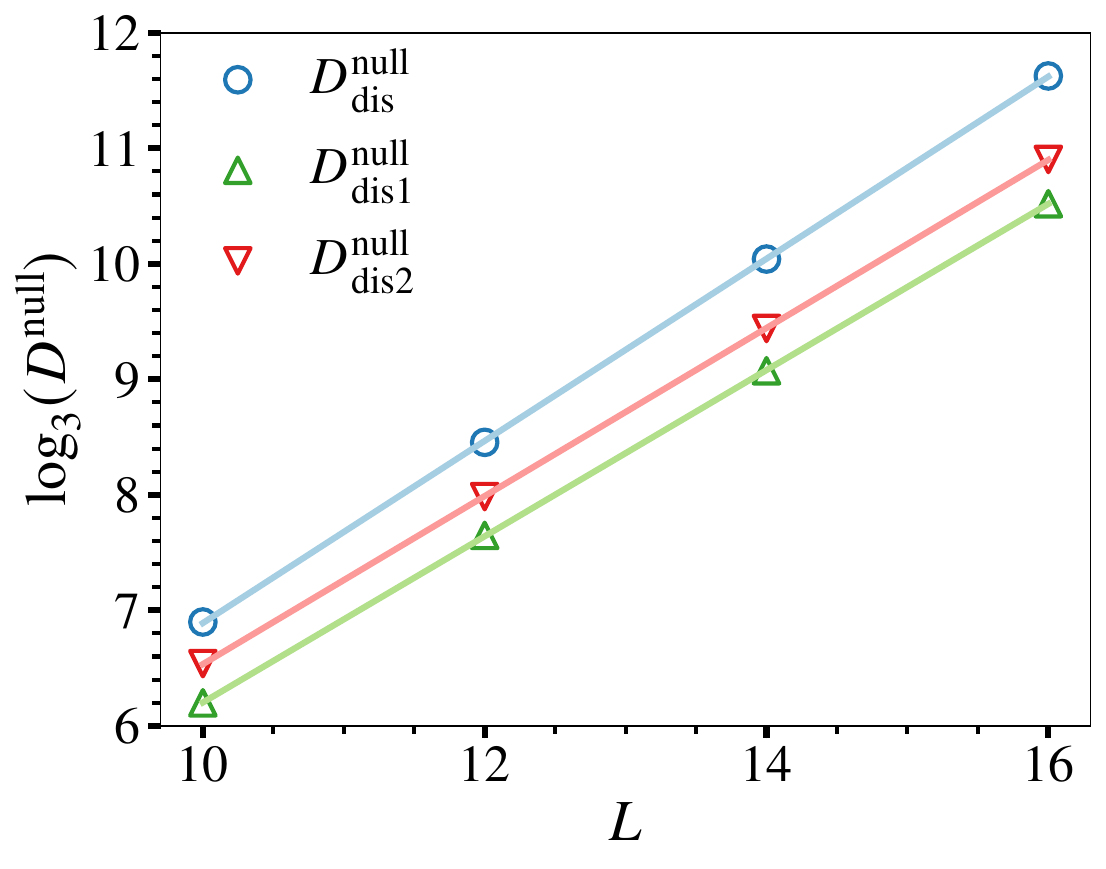}
    \end{center}
    \par
    \renewcommand{\figurename}{Fig.}
    \caption{Dimensions of the null space of the three disorder terms for different system sizes $L$. Lines indicate the linear fitting.}
    \label{Fig_Null_Scaling}
\end{figure}

\begin{figure}[thbp]
  \begin{center}
  \includegraphics[width=0.95\textwidth]{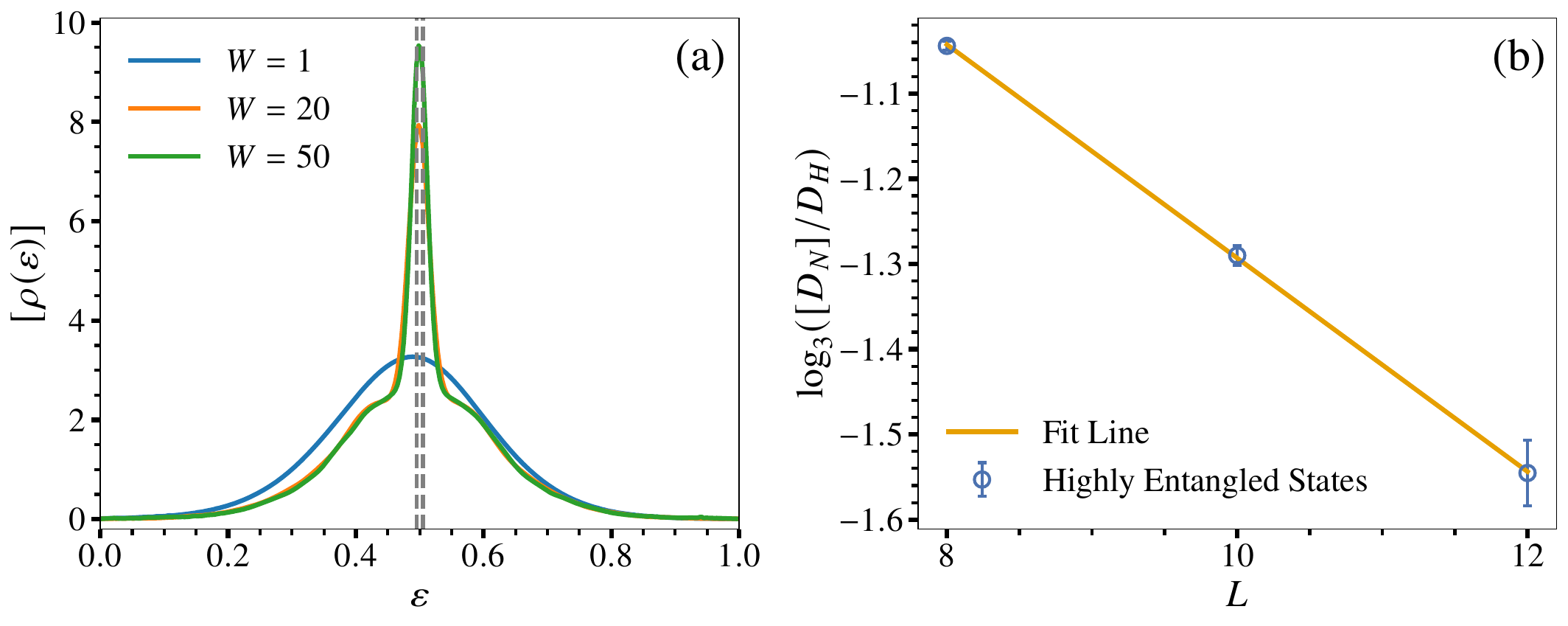}
  \end{center}

  \par
  \renewcommand{\figurename}{Fig.}
  \caption{(a) Disorder-averaged density of states for three different disorder strengths. The two vertical dashed lines represent $\varepsilon = 0.495$ and $\varepsilon = 0.505$. (b) The fraction  of the highly entangled states relative to the dimension of the $S_z = 0$ Hilbert space. The yellow line is a linear fitting, demonstrating that the thermal band becomes a measure-zero set in the thermodynamic limit.}
  \label{Fig_Dos}
\end{figure}

From Fig.~\ref{Fig_Null_Scaling}, we observe that as the system size $L$ increases, the dimension of the null space associated with all three disorder terms grows exponentially, demonstrating that these null spaces are exponentially large. This supports our claim that an exponentially large null space is critical to the phenomenon where a small fraction of highly entangled states emerges on top of low-entanglement states. Despite such scaling behavior of the null space, the fraction of the high entanglement states, $D_{N}$, with respect to the Hilbert space dimension, $D_{H}$, decreases with $L$ [see Fig.~\ref{Fig_Dos}(b)], suggesting that the highly entangled states form a zero-density subset of the Hilbert space in the thermodynamic limit.

\appendixsection{Energy gap between the low-energy manifold and its complement}

All the disorder terms discussed above possess an exponentially large low-energy manifold. When the disorder values are drawn from an interval symmetric about $0$ (i.e., $\omega_j \in [-W, W]$), each of the three disorder terms we proposed exhibits only a small and vanishing energy gap between states inside the null space and those outside it. 

The closure of this gap can be attributed to two reasons: (I) In the thermodynamic limit, spin configurations inevitably arise in which certain sites are either unaffected or only weakly influenced by the disorder, thereby eliminating any protecting gap. (II) There also exist spin configurations on which the disorder acts simultaneously on multiple sites, while the overall energy under $H_{\text{dis}}$ remains low. We refer to these as \emph{weakly disordered states}.

If the disorder values are chosen from an interval that excludes $0$, condition (I) discussed above no longer applies to any of the three disorder terms. While condition (II) still holds for $H_{\text{dis}}$ and $H_{\text{dis1}}$, it is also absent in $H_{\text{dis2}}$. This is because $H_{\text{dis2}}$ is constructed as a sum of projectors, and hence weakly disordered states do not emerge. Consequently, when the disorder window excludes $0$, the low-energy manifolds of $H_{\text{dis}}$ and $H_{\text{dis1}}$ can still be gapless with respect to their complements, whereas that of $H_{\text{dis2}}$ remains gapped even in the thermodynamic limit.

In the following sections, we examine the properties of the models when the disorder window is chosen to exclude $0$, and we fix this window to $[20, 120]$ for the subsequent discussion. In Sec.~\ref{Gapless}, we consider the case where the low-energy manifold is gapless, using $H_{\text{dis}}$ as an example, while in Sec.~\ref{Gapped} we focus on the gapped case. We further remark that a gapped low-energy manifold enables a perturbative description of the highly entangled states, which is also discussed in Sec.~\ref{Gapped}. Finally, we demonstrate that all phenomena we observe do not rely on the presence of an energy gap.

\appendixsubsection{Case 1: gapless low-energy manifold} \label{Gapless}

\begin{figure}[thb]
    \begin{center}
    \includegraphics[width=0.95\textwidth]{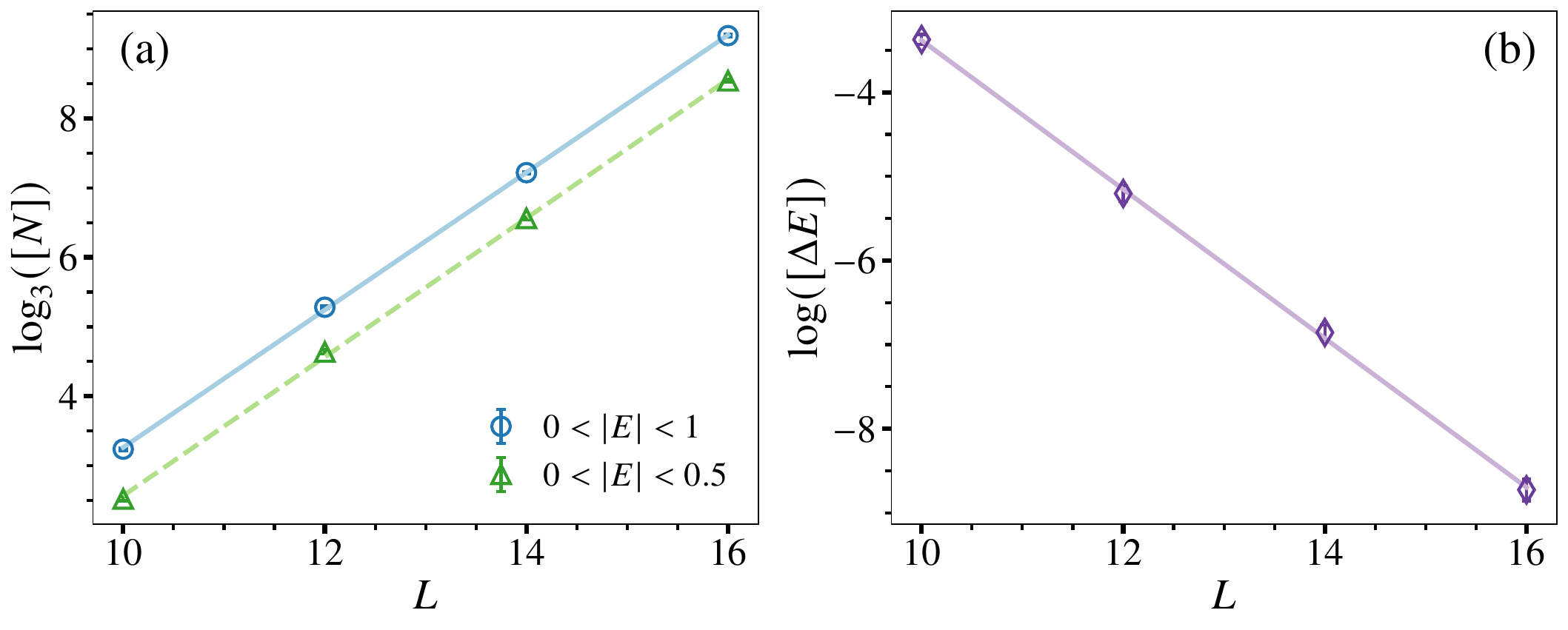}
    \end{center}
    \par
    \renewcommand{\figurename}{Fig.}
    \caption{(a) Scaling of the number for weakly disordered states against system size. All results are averaged over multiple disorder realizations. (b) The scaling of energy gap $\Delta E$ against system size.}
    \label{Weakly_disordered}
\end{figure}

We examine the properties of the weakly disordered states and the associated energy gap of $H_{\text{dis}}$. We fix the parameters as $V_0 = -0.05$, $V_1 = 0$, and $V_2 = 0.15$, while the coefficients $\omega_j$ are chosen independently and uniformly from the energy window $[20, 120]$. We then compute the number of states $N$ within an energy window around $E_{\text{dis}} = 0$, as well as the energy gap $\Delta E$ between the state closest in energy to the null space (i.e. $\Delta E = \min(|E - E_{\text{null}}|)$). The results are shown in Fig.~\ref{Weakly_disordered}.

We observe that as the system size increases, the number of weakly disordered states grows exponentially. At the same time, the energy gap between states inside and outside the null space decreases with system size, indicating that no finite gap persists in the thermodynamic limit. 

\begin{figure}[thb]
    \begin{center}
    \includegraphics[width=0.95\textwidth]{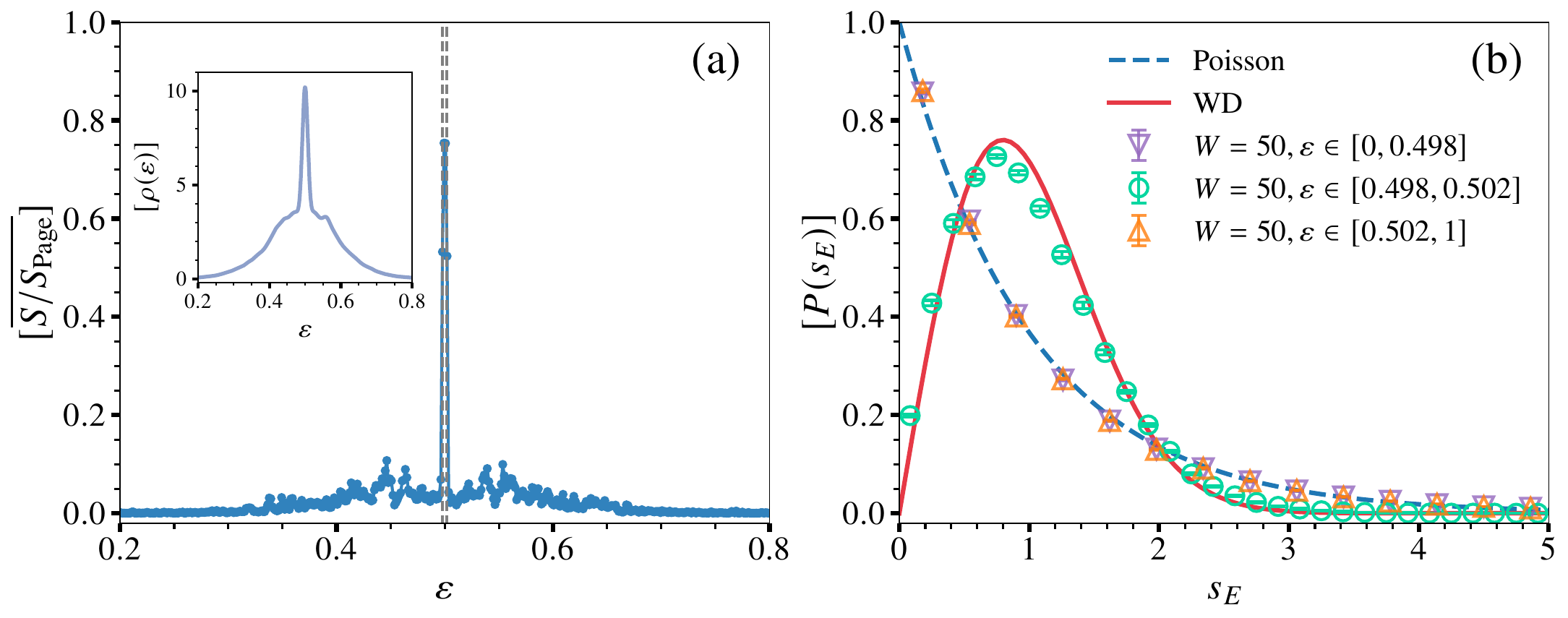}
    \end{center}
    \par
    \renewcommand{\figurename}{Fig.}
    \caption{Calculation results of the Hamiltonian $H$ using $L = 12$ and disorder window $[20, 120]$. (a) Energy-resolved disorder-averaged EE. The inset shows the DoS. The two gray dashed lines stand for $\varepsilon = 0.498$ and $\varepsilon = 0.502$. (b) Disorder-averaged level statistics of different energy intervals. The teal dots correspond to the level spacing ratio $[\bar{r}] \approx 0.517$ and the purple (orange) triangles shows $[\bar{r}] \approx 0.391$ ($[\bar{r}] \approx 0.391$). Here we only consider the central $80\%$ eigenstates across the whole spectrum.}
    \label{L12_dis}
\end{figure}

To further demonstrate that, under the disorder window $[20, 120]$, the system exhibits the same features as those obtained in the main text with a symmetric disorder window, and that these features are robust against system size, we compute the energy spectrum and the corresponding level-spacing ratio of the Hamiltonian $H_{\text{AKLT}} + H_{\text{dis}}$ for a large system size $L=12$. The results, shown in Fig.~\ref{L12_dis}, clearly reveal the emergence of highly entangled thermal states on top of a low-entanglement background, consistent with the observations presented in the main text.

\appendixsubsection{Case 2: gapped low-energy manifold} \label{Gapped}

For the disorder term $H_{\text{dis2}}$, which is constructed as a sum of projectors, weakly disordered states do not appear. Consequently, a finite energy gap persists in the thermodynamic limit when the disorder values are drawn from a window not centered at $0$. In this case, second-order perturbation theory provides an accurate description of our results.

\begin{figure}[thb]
    \begin{center}
    \includegraphics[width=0.95\textwidth]{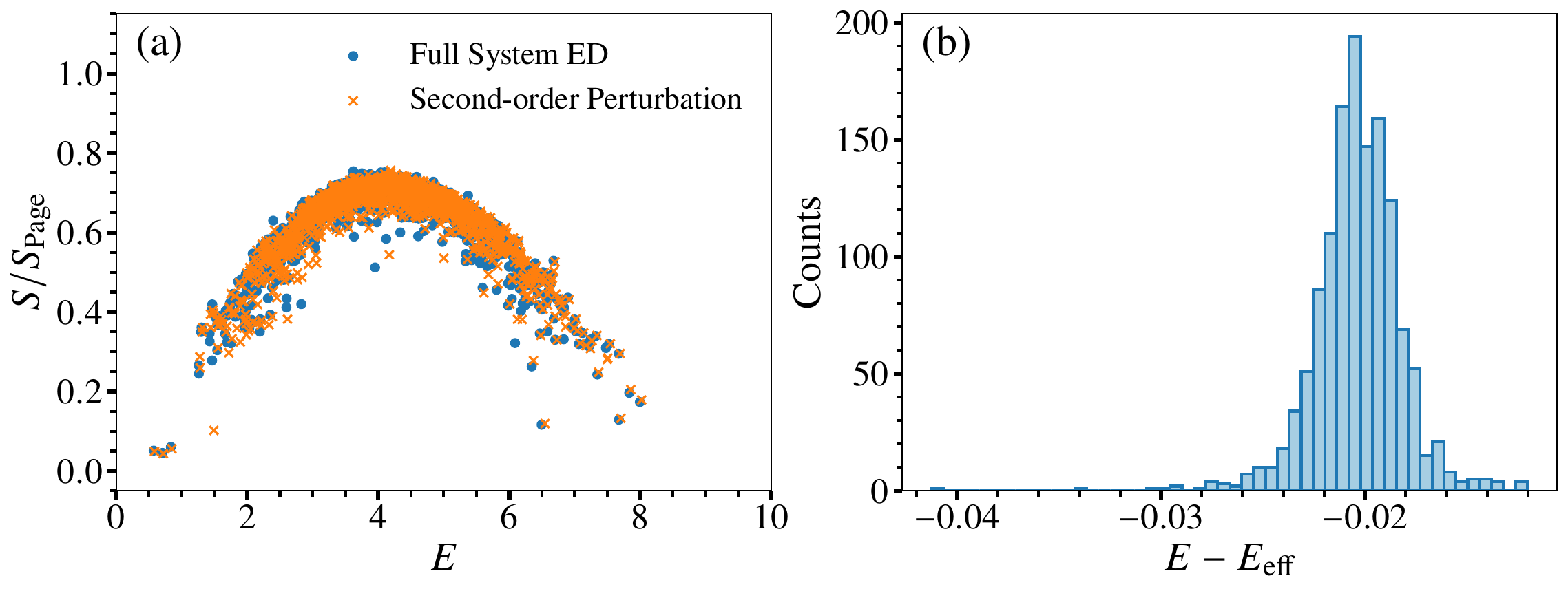}
    \end{center}
    \par
    \renewcommand{\figurename}{Fig.}
    \caption{Comparison between results calculated using the exact Hamiltonian $H_{\text{AKLT}+\text{dis2}}$ and the effective Hamiltonian $H_{\text{eff}}$ under $L=10$. Only the highly entangled part is shown.}
    \label{SW}
\end{figure}

Under this circumstance, the effective model of the low-energy manifold derived using the Schrieffer-Wolff transformation up to second order reads
\begin{equation}
    H_{\text{eff}} = PH_{\text{AKLT}}P - PH_{\text{AKLT}}Q\frac{1}{H_{\text{dis2}}}QH_{\text{AKLT}}P + O(1/W^2),
\end{equation}
where
\begin{equation}
    P \equiv \text{proj onto } \mathcal{L}, \quad Q \equiv 1 - P, \quad \mathcal{L} \equiv \{\left.|\psi\rangle \right|~ H_{\text{dis2}} |\psi\rangle = 0\}.
\end{equation}
This effective model can be used to describe the highly entangled states of the system. We again fix the disorder window to $[20, 120]$, and the corresponding results are presented in Fig.~\ref{SW}. A remarkable agreement between the exact calculations and the perturbative results is observed for the highly entangled states, indicating that the thermal states can be fully characterized by second-order perturbation theory.

It should be noted that, although the entanglement entropy results do not match exactly [see Fig.~\ref{SW}(a)]---primarily due to the extreme sensitivity of entanglement entropy to the precise form of the wavefunction---the energy spectra obtained from the exact Hamiltonian and from the effective model show excellent agreement [see Fig.~\ref{SW}(b)]. These results demonstrate that, while certain models cannot be interpreted within the perturbative framework due to the presence of weakly disordered states, there also exist models that maintain a finite energy gap between states inside and outside the null space. This indicates that the perturbative picture can successfully describe a subset of cases, though not all scenarios. 

We need to remark that, for $H_{\text{dis2}}$, when we set the disorder window to $[20, 120]$, the null space sector of the disorder term will show on the left edge of the full spectrum since the rest of spin configurations always take nonnegative energy values under the influence of the disorder term. 

\begin{figure}[thb]
    \begin{center}
    \includegraphics[width=0.95\textwidth]{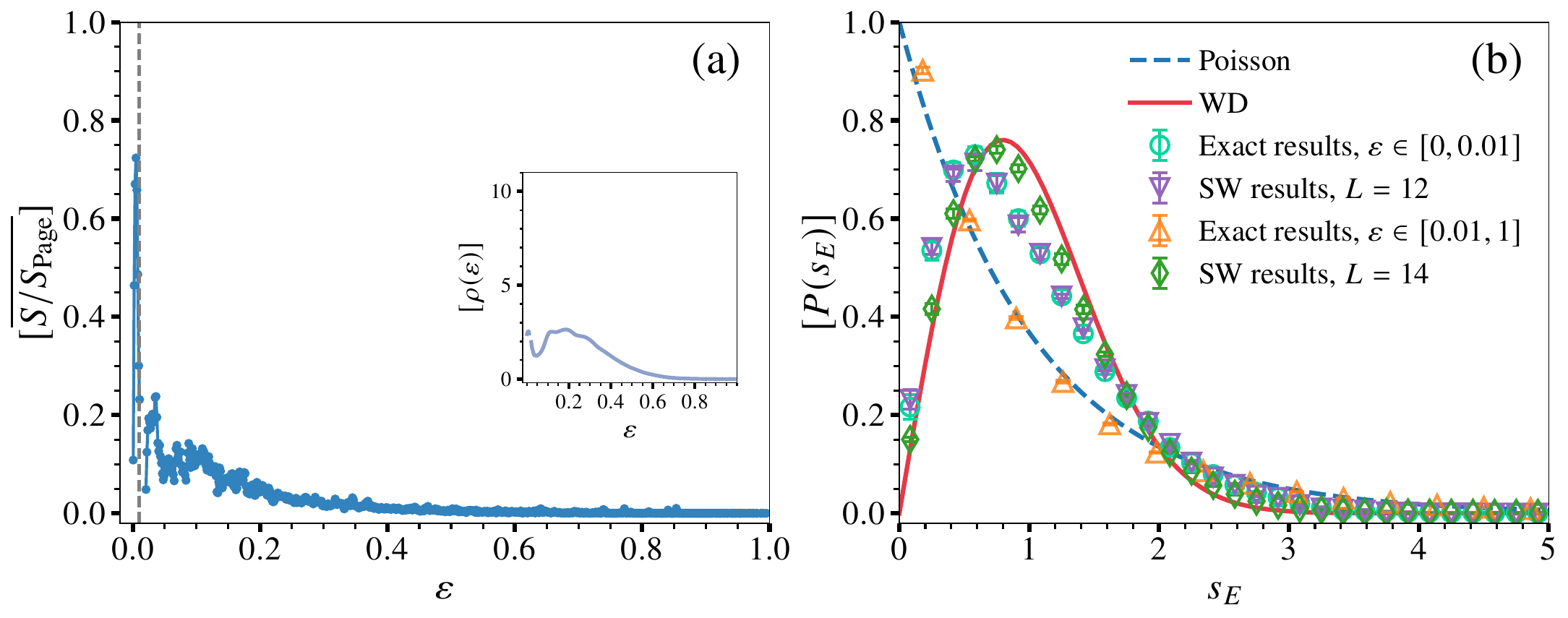}
    \end{center}
    \par
    \renewcommand{\figurename}{Fig.}
    \caption{Calculation results of the Hamiltonian $H_{\text{AKLT}+\text{dis2}}$ using $L = 12$ and disorder window $[20, 120]$ and SW calculation under $L=12$ and $L=14$. (a) Energy-resolved disorder-averaged EE. The inset shows the DoS. The gray dashed line represents $\varepsilon = 0.01$. (b) Disorder-averaged level statistics of different energy intervals. The teal dots correspond to the level spacing ratio $[\bar{r}] \approx 0.481$ and the orange triangles show $[\bar{r}] \approx 0.390$. The green diamonds stand for the level spacing ratio $[\bar{r}] \approx 0.521$. Here we only consider the central $80\%$ eigenstates across the whole spectrum.}
    \label{L12_dis2}
\end{figure}

We also examine the energy spectrum and level statistics of $H_{\text{AKLT}+\text{dis2}}$ under disorder window $[20, 120]$ for $L=12$, and perform the SW approximation for $L=12$ and $L=14$. The results are presented in Fig.~\ref{L12_dis2}. Since all states outside the null space of $H_{\text{dis2}}$ acquire nonnegative energies, the highly entangled states appear at the left edge of the spectrum [see Fig.~\ref{L12_dis2}(a)]. For $L=12$, the limited size of the null space prevents the level statistics of the highly entangled states from approaching the WD distribution. This issue disappears when the system is enlarged to $L=14$, where the level statistics given by SW approximation of the highly entangled states approach the WD distribution, and the corresponding level spacing ratio converges to the GOE value. These results demonstrate that for $H_{\text{AKLT}+\text{dis2}}$, even under disorder window $[20, 120]$ and large system size, there emerge some highly entangled thermal states embedded in a background of low-entanglement states. 

Taken together, our discussion shows that the low-energy manifold of the disorder term can be either gapless or gapped with respect to its complement. In both scenarios, however, the phenomenon of highly entangled thermal states emerging on a background of weakly entangled states persists and is robust against varying system size.

\appendixsection{Additional results for the disordered AKLT model in the main text}

In the following, we present additional results for the disordered AKLT model introduced in Eq.~(\ref{Model_AKLT}) of the main text. Specifically, we show the density of states, provide further details on entanglement entropy growth following quench dynamics, present additional results on total fidelity, and include calculations for a chain of length $L=12$.

\appendixsubsection{Density of states}

In Fig.~\ref{Fig_Dos}(a), we present the density of states (DoS) for three different disorder realizations. Under weak disorder, the DoS remains nearly Gaussian. Under large disorder strength, a peak emerges around the center of the spectrum, showing great correspondence with the energy window where the highly entangled states show up. 

Here, we also demonstrate the scaling behavior of the fraction of the high entanglement states, $D_{N}$, with respect to the Hilbert space dimension, $D_{H}$ [see Fig.~\ref{Fig_Dos}(b)]. Note that the high entanglement states are selected if their EE exceeds $0.4S_{\text{Page}}$. The percentage of the highly entangled states decreases as the size of the system grows. This indicates that the highly entangled states will only form a measure-zero subset of the total Hilbert space under the thermodynamic limit.

We note that, although a similar phenomenon---namely the emergence of highly entangled states within a low-entanglement background---was also reported in Ref.~\cite{PhysRevB.109.125127}, our case is distinct in two respects: (i) the low-entanglement states in our system form a uniform non-thermal background, and (ii) the highly entangled states constitute a measure-zero subset of the total Hilbert space.

\appendixsubsection{Details on EE growth after quantum quench}

\begin{figure}[thbp]
  \begin{center}
  \includegraphics[width=0.95\textwidth]{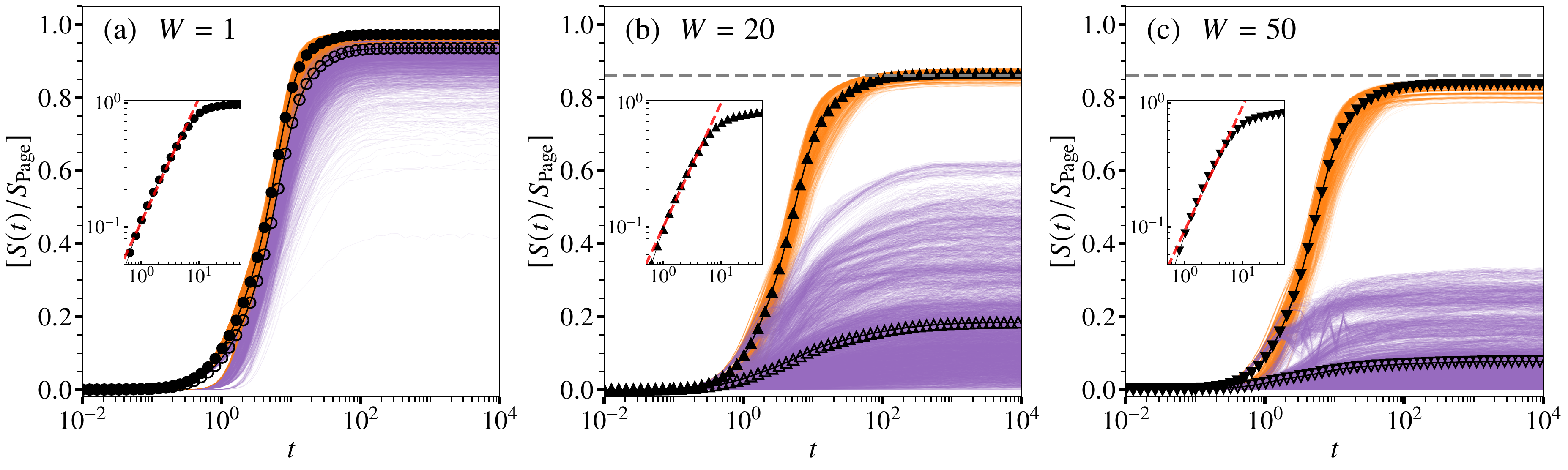}
  \end{center}

  \par
  \renewcommand{\figurename}{Fig.}
  \caption{Disorder-averaged time evolution of EE under different disorder strength. From (a) to (c), the disorder strength takes $W=1$, $W=20$ and $W=50$, respectively. In all three panels, the black filled dotted line and the unfilled dotted line are the same as the lines of corresponding disorder strength in Fig.~\ref{Fig_highEE}(b) in the main text, but with a finer time resolution. The orange (purple) lines indicate the disorder-averaged $S(t)$ of each product state inside (outside) the null space of $H_{\text{dis}}$. The red line in the inset shows the linear fit $S_{\text{fit}}(t) = k_1 t$ of the initial growth of $S(t)$, with $k_1 \approx 0.541$, $k_2 \approx 0.487$ and $k_3 \approx 0.463$ corresponding to fitting results from (a) to (c). }
  \label{Fig_Quench_detail}
\end{figure}

In this section, we present some further details on the quench dynamics of the disorder-averaged EE $[{S(t) / S_{\text{Page}}}]$. Specifically, we examine  $[{S(t) / S_{\text{Page}}}]$ from different initial states, supplementing Fig.~\ref{Fig_highEE}(e) in the main text, which presented the state- and disorder-averaged dynamics $[\overline{S(t) / S_{\text{Page}}}]$. We find that, as shown in Fig.~\ref{Fig_Quench_detail}(a) with weak disorder, the $[{S(t) / S_{\text{Page}}}]$ for states inside (orange lines) and outside (purple lines) the null space of $H_{\text{dis}}$ are very similar. This again demonstrates that, under weak disorder, all states in this system go to thermalization. 

However, as disorder strength increases, a pronounced divergence emerges in the dynamics between product states within and outside the null space of $H_{\text{dis}}$. Figure~\ref{Fig_Quench_detail}(b) and (c) reveal a stark divergence in the dynamics of product states inside (orange lines) and outside (purple lines) the null space of $H_{\text{dis}}$. With increasing disorder strength, initial states within the null space (orange lines) evolve to high-entanglement states, with trajectories of $[{S(t) / S_{\text{Page}}}]$ remaining closely concentrated, consistent with thermalization. In contrast, states outside the null space (purple lines) exhibit suppressed entanglement growth, with $S(t)$ saturating at significantly lower values. This behavior further confirms that the thermalization dynamics are strongly associated with the null space sector of $H_{\text{dis}}$. The insets in Fig.~\ref{Fig_Quench_detail}(a)-(c) show the linear fit of the initial growth of $[{S(t) / S_{\text{Page}}}]$. The excellent agreement between the fits and the early-time linear increase of EE further reinforces the notion of highly chaotic, ballistic entanglement propagation before saturation.

\appendixsubsection{Total fidelity for other disorder realization}

\begin{figure}[tbh]
  \begin{center}
  \includegraphics[width=0.95\textwidth]{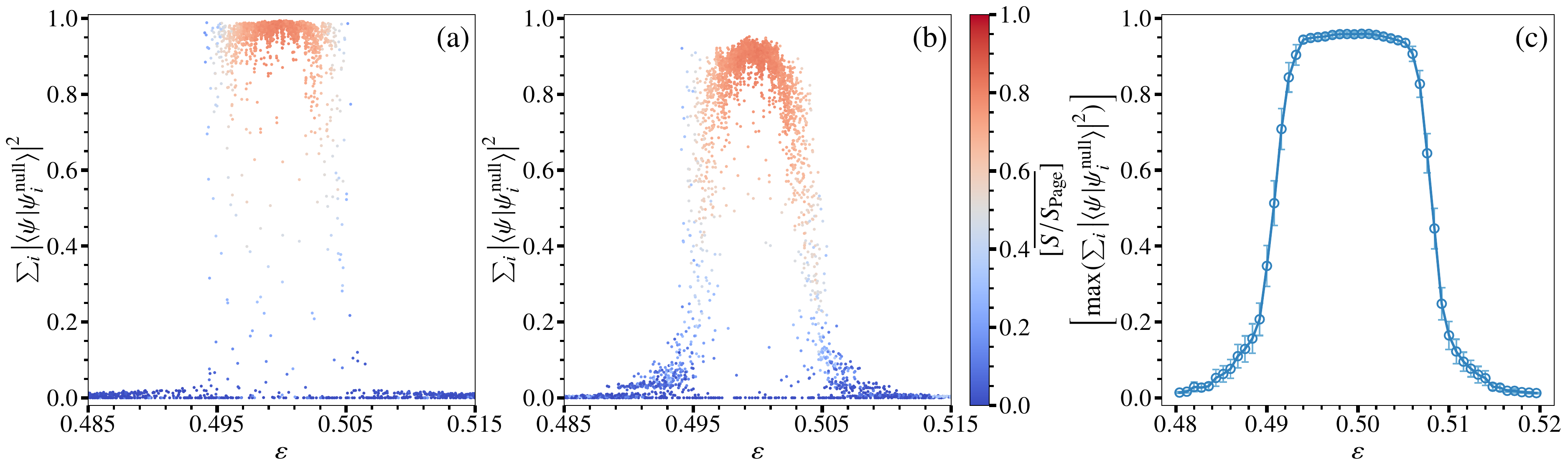}
  \end{center}
  \par
  \renewcommand{\figurename}{Fig.}
  \caption{(a) and (b) Total fidelity calculated the same as Fig. \ref{Fig_Overlap} but for different disorder realizations. (c) Energy-resolved disorder-averaged maximum fidelity in energy windows calculated under $L=10$ and $W=50$.}
  \label{Fig_Overlap_appen}
\end{figure}

Here we present further calculation results on the total fidelity but using different disorder realizations from the main text. We can clearly find from Fig.~\ref{Fig_Overlap_appen}(a), (b) that, although different disorder configurations will lead to different results in total fidelity in a quantitative way, the overall character of the total fidelity stays the same. Among different disorder realizations, most eigenstates within the energy window $\varepsilon \in [0.495, 0.505]$ show both high EE and high total fidelity, while states outside this interval only display limited fidelity. This confirms that the high entangled thermal states are supported by the null space of $H_{\text{dis}}$ and corroborates that this conclusion is robust among different disorder realizations. 

To further confirm that the thermal states include small admixtures from outside the null space, we compute the disorder-averaged maximum total fidelity within energy windows, defined as $\left[\max\left(\sum_i |\langle \psi | \psi_{i}^{\text{null}} \rangle |^2\right)\right]$, where $\max(\cdot)$ is taken over all states within each interval. As shown in Fig.~\ref{Fig_Overlap_appen}(c), this quantity reaches relatively high values near the center of the energy spectrum but remains below unity. This indicates that, although the thermal states are predominantly composed of components within the null space of $H_{\text{dis}}$, they still contain small contributions from outside of it. This distinguishes the thermal states from the disorder-annihilated states.

\begin{figure}[htb]
  \begin{center}
  \includegraphics[width=0.95\textwidth]{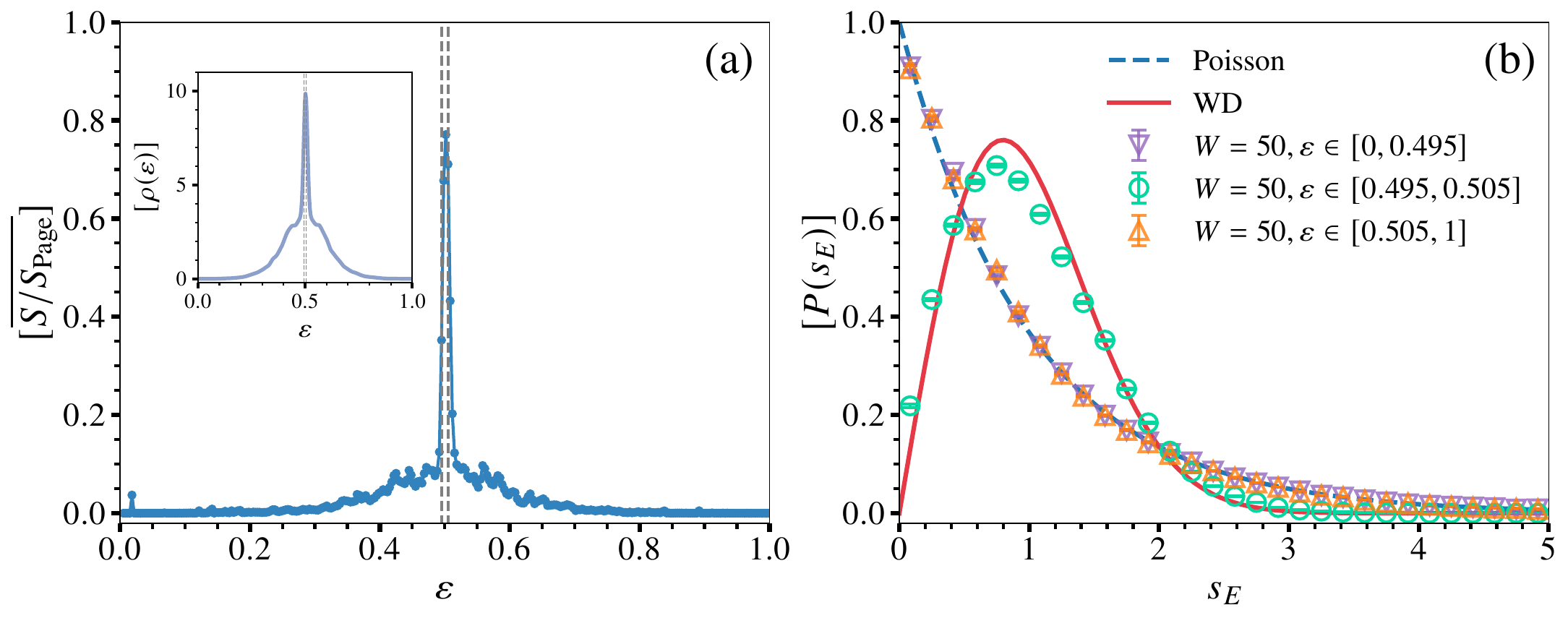}
  \end{center}

  \par
  \renewcommand{\figurename}{Fig.}
  \caption{Calculation results using $L = 12$ and $W = 50$. (a) Energy-resolved disorder-averaged EE. The inset shows the DoS. The two gray dashed lines stand for $\varepsilon = 0.495$ and $\varepsilon = 0.505$. (b) Disorder-averaged level statistics of different energy intervals. The teal dots correspond to the level spacing ratio $[\bar{r}] \approx 0.515$ and the purple (orange) triangles shows $[\bar{r}] \approx 0.393$ ($[\bar{r}] \approx 0.394$). Here we only consider the central $80\%$ eigenstates across the whole spectrum.}
  \label{Fig_L12}
\end{figure}

\appendixsubsection{Results of $L = 12$}
We also present the calculation results under $L = 12$ and within the $S^z_{\text{tot}}=0$ sector. As is shown in Fig.~\ref{Fig_L12}(a), we can see that, for $L = 12$, under strong enough disorder, the energy-resolved EE sees a distinct peak around a narrow energy window $\varepsilon \in [0.495, 0.505]$, where the DoS shows a peak as well. Both are in line with our previous results on $L = 10$.

The disorder-averaged level statistics are also consistent with those acquired for $L=10$. On the one hand, for eigenstates within the energy window where the highly entangled states appear, a clear match with WD distribution can be seen. The corresponding level spacing ratio is also close to  {$r_{\text{GOE}}$}, indicating quantum chaotic behaviors. On the other hand, for eigenstates outside this energy window, the level spacing ratio typically follows a Poisson distribution, with level spacing ratios approaching {$r_{\text{Poisson}}$}, both are features of a non-ergodic system. 

\end{appendix}
\end{document}